\documentclass[12pt]{article}
\usepackage{amssymb, amsmath, amscd, mathtools}
\usepackage{graphicx}
\usepackage[utf8]{inputenc}
\usepackage[T1]{fontenc}

\usepackage{epsfig}
\usepackage{color}
\usepackage{subfig, color}
\usepackage{resizegather}

\usepackage[backref=page]{hyperref}

\usepackage{algorithm}
\usepackage{algorithmic}

\usepackage{etoolbox}
\patchcmd{\thebibliography}{\section*{\refname}}{}{}{}
\usepackage[bottom]{footmisc}

\begin{document}

\title{Graver Bases via Quantum Annealing\\with application to non-linear integer programs}
\author{Hedayat Alghassi\footnote{halghassi@cmu.edu},\,  Raouf Dridi\footnote{rdridi@andrew.cmu.edu},\, Sridhar Tayur\footnote{stayur@cmu.edu} \\
\small CMU Quantum Computing Group\\
\small Tepper School of Business, Carnegie Mellon University, Pittsburgh, PA 15213\\}
{%\small $\{$halghassi, rdridi, stayur$\}$@cmu.edu}}

\maketitle
\begin{abstract}
We propose a novel hybrid quantum-classical approach to calculate Graver bases, which have the potential to solve a variety of hard linear and non-linear integer programs, as they form a test set (optimality certificate) with very appealing properties. The calculation of Graver bases is exponentially hard (in general) on classical computers, so they not used for solving practical problems on commercial solvers. With a quantum annealer, however, it may be a viable approach to use them. We test two hybrid quantum-classical algorithms (on D-Wave)--one for computing Graver basis and a second for optimizing non-linear integer programs that utilize Graver bases--to understand the strengths and limitations of the practical quantum annealers available today. Our experiments suggest that with a modest increase in coupler precision--along with near-term improvements in  the number of qubits and connectivity (density of hardware graph) that are expected--the ability to outperform classical best-in-class algorithms is within reach, with respect to non-linear integer optimization.
\end{abstract}

\textbf{Keywords}: Graver bases, quantum annealing, test sets, non-linear integer optimization, computational testing.
%\maketitle

\newpage
\tableofcontents 
\newpage

\section{Introduction}
\label{sec:introduction}

This paper explores the capability of an available quantum annealer for solving non-linear integer optimization problems (that arise in engineering and business applications) using test sets. Of course, we do not expect current quantum annealers (D-Wave 2000Q, in our setting) to be competitive against classical computing algorithms that have decades of innovations and implementation improvements. Our motivation in this work is more of a proof-of-concept: to check how viable a hybrid quantum-classical approach is and understand what it would take to surpass classical methods.   

~~\\
The D-Wave quantum annealer solves Quadratic Unconstrained Binary Optimization (QUBO) problems of the form \cite{berwald_mathematics_2018}:
\begin{equation}\label{eq:qubo}
% \begin{align}
\begin{array}{*{20}{c}}
{min}&{{x^T}Qx}&{\mbox{such\, that}}&{x \in {{\left\{ {0,1} \right\}}^n}}.\\
\end{array}
% \end{align}
\end{equation}
In this paper, we develop a hybrid quantum-classical algorithm that calculates a test set called Graver bases, on such a quantum annealer, and then we use it to optimize nonlinear integer programming problems.  
Because Graver bases \cite{graver_foundations_1975} have applications in areas other than optimization, studying how our algorithm acquires Graver bases of integer matrices of various kinds is of interest in itself.
We test our approach on D-Wave and report our findings.
%Later we focus on using these Graver basis or truncated version of them as test sets and solve nonlinear integer problems using augmentations from initial feasible solutions.
\subsection{Illustrative Example}\label{subsec:illustrative}
The following example walks through some of the key steps of our algorithms (and reviews the definition of Graver bases). Those steps will be described in detail in later sections. 
 Consider the following integer optimization problem with a non-linear objective function:
\begin{equation}
\left\{ {\begin{array}{*{20}{c}}
  {\begin{array}{*{20}{c}}
  {\min }&{f(x) = \sum\limits_{i = 1}^4 {\left| {{x_i} - 5} \right|} } 
\end{array}\begin{array}{*{20}{c}}
  ,&{{x_i} \geqslant 0}&,&{{x_i} \in \mathbb{Z}} 
\end{array}} \\ 
  {\left[ {\begin{array}{*{20}{c}}
  1&1&1&1 \\ 
  1&5&{10}&{25} 
\end{array}} \right]\left[ {\begin{array}{*{20}{c}}
  {{x_1}} \\ 
  {{x_2}} \\ 
  {{x_3}} \\ 
  {{x_4}} 
\end{array}} \right] = \left[ {\begin{array}{*{20}{c}}
  {21} \\ 
  {156} 
\end{array}} \right]} 
\end{array}} \right.
\nonumber
\end{equation}
Let $A$ denote the $2 \times 4$ integer matrix above; so we write the above constraint as $Ax=b$. By definition, the Graver basis $\mathcal G(A)$  of the matrix $A$ is a minimal set of elements of the kernel $\mathrm{ker}_\mathbb{Z}A$ with respect to the partial ordering 
$$
x \sqsubseteq y  \mbox{ when }   {x_i}{y_i} \geqslant 0 \mbox{ (the two vectors $x$ and $y$ lie on the same orthant) and } \left| {{x_i}} \right| \leqslant \left| {{y_i}} \right|.
$$
The steps of our hybrid quantum-classical algorithm for computing Graver bases are:
\begin{itemize}
\item We map the problem of finding the kernel of $A$ into a QUBO of the form (\ref{eq:qubo}). This is the quantum part of the hybrid approach.
\item The  elements of $\mathcal G(A)$ can be obtained from the degenerate solutions of the given QUBO. The way such elements are obtained (filtered from kernel elements, followed by some post-processing) constitutes the classical part of our hybrid approach.
\end{itemize}
In our example, the Graver basis elements are the column vectors of the matrix:
\begin{equation}
\mathcal G(A) = \left[ {\begin{array}{*{20}{c}}
0&5&5&5&5\\
3&{ - 9}&{ - 6}&{ - 3}&0\\
{ - 4}&4&0&{ - 4}&{ - 8}\\
1&0&1&2&3
\end{array}} \right]
 = \left[ {\begin{array}{*{20}{c}}
  {{g_1}}&{{g_2}}&{{g_3}}&{{g_4}}&{{g_5}} 
\end{array}} \right]
\nonumber
\end{equation}
This was obtained using only one call to the D-Wave quantum annealer (details in \ref{example1}) and some classical post-processing.  (Typically, more than one call is required. We detail this in a later section.) Once the Graver basis is computed, the optimal solution can be obtained by augmenting, beginning from any feasible solution.  To solve the optimization problem, our second hybrid quantum-classical algorithm is as follows:
\begin{itemize}
\item We use the quantum annealer (once more) to solve the constraint $Ax=b$ (using QUBO in Equation (\ref{eq:axbqubo})) and obtain a feasible solution.  
(Actually, a call of D-Wave usually gives many solutions, up to 10000. We discuss how many of these are unique, sometimes more than 6000, and which one(s)to use, in later subsections).
\item With the Graver basis and a feasible solution at hand, we can start the \textit{augmentation} procedure (which is classical portion of our second algorithm). 
\end{itemize}

~~\\
Let us illustrate the augmentation step.
An initial feasible solution obtained is
${x^0} = {\left[ {\begin{array}{*{20}{c}}
  1&15&3&2
\end{array}} \right]^T}$ with    $f(x^0) = 19$.  Adding the vector 
${-g_1} = {\left[ {\begin{array}{*{20}{c}}
  0&-3&4&-1 
\end{array}} \right]^T}$ to the previous value gives 
${x^1} = {\left[ {\begin{array}{*{20}{c}}
  1&12&7&1 
\end{array}} \right]^T}$,
which has a lower cost value of $f(x^1) = 17$. Similarly, adding 
${g_2} = {\left[ {\begin{array}{*{20}{c}}
  5&-9&4&0 
\end{array}} \right]^T}$ to the previous value yields
${x^2} = {\left[ {\begin{array}{*{20}{c}}
  6&3&11&1 
\end{array}} \right]^T}$,
which has a lower cost value of $f(x^2) = 13$. Similar calculations are done consecutively with the vectors $g_5, -g_3$ and $-g_1$ again, which finally results in  
${x^5} = {\left[ {\begin{array}{*{20}{c}}
  6&6&7&2
\end{array}} \right]^T}$,
which has the cost value   $f(x^5) = 7$. Adding any of the Graver basis elements again to $x^5$ results in vectors that all have a cost higher than the current minimum $7$ (inside the feasible region~$x \geqslant 0$); therefore, the augmentation procedure stops with the optimal solution ${x^*} = {x^5}$.

~~\\
% We want to mention a few features of our procedure.
While the objective function can be linearized, it would require increasing the number of variables, an important consideration when we have limited number of qubits and connectivity. Furthermore, other non-linear objective functions that may not be easily linearizable can also be handled with the same Graver basis and feasible solutions, making our procedure more general purpose.

\subsection{Features and Contributions of our Approach}
We list below some key features followed by key contributions of our approach.
\subsubsection{Features}

\begin{itemize}
\item  Graver basis is an optimality certificate, and the calculation of elements depends only on the constraints. Thus, the objective function can even be a comparison oracle. The objective function appears only in the augmentation step where it is called (queried). Several different objective functions (for the same constraints) can be studied classically once the Graver basis has been computed.
\item In the special case of a quadratic objective function,  a naive approach of combining the objective function and the constraints through a Lagrangian multiplier into a single QUBO may appear attractive. However, searching for the right Lagrangian multiplier is non-trivial. We avoid this complexity in our approach as we have decoupled the objective function from the constraints.
\item In the case of an objective function that is a polynomial function, there is no need for any reduction to QUBO and adding extra variables, different from other methods that utilize a quantum annealer.  

\item At each anneal (read), the algorithm starts out fresh from the uniform superposition 
% $$
% |\hat 0\rangle = \frac{1}{2^n}\sum _{i\in \mathbb Z_2^n} |i\rangle
% $$ 
(superposition of the states of the computational basis of ${\mathbb C^2}^{\otimes n}$). A single call can have up to $10000$ reads. Thus, it can reach different ground states with good probabilities (i.e., be able to extract different kernel elements, $\mathrm{ker}_{\mathbb Z}(A) \subset \mathbb Z_2^n$,), and so within a few calls, we can expect to find all of them. 

\end{itemize}

 \subsubsection{Contributions}
\begin{itemize}
\item We provide two hybrid quantum classical algorithms: (1) to calculate the Graver basis of any general integer matrix, and (2) to solve a non-linear integer programming problem using the Graver basis. By separating the objective function and the constraints equation's ($Ax = b$) right hand side ($b$) from the calculation of the test set (Graver basis), our approach can tackle many non-linear objectives, cost functions, only via an oracle call, and also help in solving deterministic and stochastic programs (with varying right hand sides $b$ or objective function parameters) in a systematic manner.

\item For the Graver calculation algorithm, we develop a (classical) systematic post-processing strategy that transforms the quantum annealer's near-optimal solutions into optimal solutions, which might be an useful technique in other situations.

\item For iterative integer to binary transformation in these algorithms, we  develop an adaptive strategy that moves (a) the middle point and (b) the encoding length of integer variables based on the suboptimal solutions obtained in the previous iteration. This strategy, which optimizes resources (qubits), can be used in any integer to binary encoding that is solved iteratively by a quantum annealer.

\item We test both Graver and non-linear optimization algorithms,  on the D-Wave 2000Q quantum annealing processor and discuss the implementation details, strengths, and shortcomings. 

\item Unlike the typical application of a quantum annealer that uses the many reads to increase the confidence of a desired solution, we use the many reads to sample the many unique solutions. This re-purposing  makes our approach particularly suitable to tackle hard integer optimization problems.

\item   Our experiments were designed to elucidate what is preventing the quantum annealer from being competitive today, and what enhancements are needed (and why) to surpass classical methods. What we find is that even if we have a processor with the same number of qubits and connectivity structure, with a \textit{higher coupler precision}, some classes of nonlinear integer problems can be solved comparably to best-in-class classical solvers, e.g., Gurobi Optimizer (latest version, 8.0). Alternatively, with a new processor with a higher number of qubits and better connectivity that has already been announced, even without an increase in coupler precision, certain instances can be solved better than classical methods. Indeed, doubling the coupler precision (on top of the announced increases in qubits and connectivity) would lead to surpassing classical methods for many problems of practical importance.

\end{itemize}
 \subsection{Outline}
 
We are bringing together (a) the calculation of Graver bases using quantum annealing and~(b)~solving integer optimization through Graver bases with  feasible solutions obtained via quantum annealing. Section \ref{sec:background} introduces nonlinear integer programs and Graver bases. A pre-requisite to acquiring the Graver basis is to find the kernel of a matrix. Section \ref{sec:graverbasis} develops a hybrid quantum classical algorithm for calculation of Graver basis of an integer matrix and presents computational results.  Section \ref{sec:nonlinearopt} develops the quantum classical algorithm for non-linear integer programs and presents numerical results. Section \ref{sec:conclusions} summarizes our findings. To make this document self-contained, our appendices briefly review (A) computing Graver bases classically and (B) quantum annealing with D-Wave processors.

\section{Background: Graver Bases and its Use in Non-linear Integer Optimization}\label{sec:background}

Let ${f:{\mathbb{R}^n} \to \mathbb{R}}$  be a real-valued function.
We want to solve the general non-linear integer optimization problem:
\begin{equation}\label{eq:gennonlin}
{(IP)_{A,b,l,u,f}} : \left\{ {\begin{array}{*{20}{l}}
  {\begin{array}{*{20}{c}}
  {\min }&{f(x)},&& 
\end{array}} \\ 
  {\begin{array}{*{20}{c}}
  {Ax = b},&&{l \leqslant x \leqslant u},&&{x,l,u \in {\mathbb{Z}^n}} 
\end{array}} \\ 
  {\begin{array}{*{20}{c}}
  {A \in {\mathbb{Z}^{m \times n}}},&&{b \in {\mathbb{Z}^m}} 
\end{array}} 
\end{array}} \right.
\end{equation}
One approach to solving such problem is to use an augmentation procedure: start from an initial feasible solution (which itself can be hard to find) and take an improvement step--{\it augmentation} - until one reaches the optimal solution. Augmentation procedures such as these need {\it test sets} or {\it optimality certificates}:  so it either declares the optimality of the current feasible solution or provides direction(s) towards better solution(s). Note that it does not matter from which feasible solution one begins, nor the sequence of improving steps taken: the final stop is an optimal solution. (It is not hard to see that computing a test set can be very expensive classically. We want to study whether quantum annealing can be competitive.)
\paragraph{Definition 1.} 
A set  $\mathcal{S} \in {\mathbb{Z}^n}$ is called a test set or optimality certificate if for every non-optimal but feasible solution ${x_0}$ there exists  $t \in \mathcal{S}$ and  $\lambda  \in {\mathbb{Z}_ + }$ such that  ${x_0} + \lambda t$ is feasible and $f\left( {{x_0} + \lambda t} \right) < f\left( {{x_0}} \right)$. The vector $t$ (or $\lambda t$)  is called the  \textit{improving} or \textit{augmenting} direction.\\
\\
If the optimality certificate is given, any initial feasible solution ${x_0}$ can be augmented to the optimal solution.  If $\mathcal{S}$ is finite, one can enumerate over all $t \in \mathcal{S}$ and check if it is augmenting (improving). If $\mathcal{S}$ is not practically finite, or if all elements $t \in \mathcal{S}$ are not available in advance, it is still practically enough to find a subset of $\mathcal{S}$ that is feasible and augmenting.  

\begin{figure}[H]
\centering
\includegraphics[width=3.5cm]{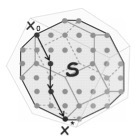}
\caption{Augmenting from initial to optimal solution \cite{onn_nonlinear_2010}.}
\label{fig:augmentation}
\end{figure}

~~\\
In the next section, we discuss the Graver basis of an integer matrix   $A \in {\mathbb{Z}^{m \times n}}$ which is known to be an  {optimality certificate}. 

\subsection{Mathematics of Graver Bases}
%We briefly review the basic notions and list important properties that we have employed 
First, on the set ${\mathbb{R}^n}$,  we define
the following partial order:

\paragraph{Definition 2.} 
Let  $x,y \in {\mathbb{R}^n}$.
We say $x$  is {\it conformal} to $y$, 
written  $x \sqsubseteq y$, when   ${x_i}{y_i} \geqslant 0$ ($x$ and $y$ lie on the same orthant) and $\left| {{x_i}} \right| \leqslant \left| {{y_i}} \right|$  for $i = 1,...,n$. 
Additionally, a sum $u = \sum\limits_i {{v_i}}$ is called {\it conformal}, if ${v_i} \sqsubseteq u$ for all $i$ ($u$ \textit{majorizes} all $v_i$).  
%therefor all $v_i$ are in the same orthant.

% We equip  ${\mathcal{L}^*}(A)$ with 
% a partial order:   $x \sqsubseteq y$ \footnote{I dont like conformal- The word conformal relates to Riemann surfraces etc}    when   ${x_i}{y_i} \geqslant 0$ ($x$ and $y$ are on the same orthant) and $\left| {{x_i}} \right| \leqslant \left| {{y_i}} \right|$  for $i = 1,...,n$.\\
% A sum $u = \sum\limits_i {{v_i}}$ is conformal, since ${v_i} \sqsubseteq u$ for all $i = 1,...,n$. 

~~\\
Suppose $A$ is a matrix in $ {\mathbb{Z}^{m \times n}}$. Define :   
\begin{equation}\label{eq:kernel}
{\mathcal{L}^*}(A) = \left\{ {x\left| {\begin{array}{*{20}{c}}
  {Ax = {\mathbf{0}},}&{x \in {\mathbb{Z}^n}\begin{array}{*{20}{c}}
  ,&{A \in {\mathbb{Z}^{m \times n}}} 
\end{array}} 
\end{array}} \right.} \right\}\backslash \left\{ {\mathbf{0}} \right\}.
\end{equation}
The notion of Graver basis was first introduced in \cite{graver_foundations_1975} for integer linear programs (ILP):
\paragraph{Definition 3.} 
The Graver basis  of   integer matrix $A$  is defined to be the finite set of $\sqsubseteq $ minimal elements (\textit{indecomposable} elements) in the lattice  ${\mathcal{L}^*}(A)$. We denote 
by $\mathcal{G}(A) \subset {\mathbb{Z}^n}$ the Graver basis of $A$. 

~~\\
The following proposition summarizes the properties of Graver bases that are relevant to our setting. 
%(proofs in \cite{graver_foundations_1975,sturmfels_grobner_1995, onn_nonlinear_2010} and \cite{DBLP:journals/jct/CookFS86, 
%sebo_hilbert_1990}). {\color{red} Hedayat: Please add /edit the refs here! important !}
\paragraph{Proposition 1.} 
The following statements are true: 
\begin{itemize}
\item[(i)] Every vector $x$  in the lattice   ${\mathcal{L}^*}(A)$   is a conformal sum of the Graver basis elements.
\item[(ii)]  Every vector $x$  in the lattice   ${\mathcal{L}^*}(A)$   can be written as 
$
{x = \sum\limits_{i = 1}^t {{\lambda _i}} {g_i}}$ for some $ \lambda _i \in {\mathbb{Z}_ + }$ and $ 
   g_i \in \mathcal{G}(A) $. The upper bound on  number of Graver basis elements required ($t$) (called \textit{integer Caratheodory number}) is $\left( {2n - 2} \right)$.
   
\item[(iii)] A Graver basis is a test set (optimality certificate) for ILP. That is a point ${x^*}$ is optimal for the optimization problem ${(IP)_{A,b,l,u,f}}$, if and only if there are no ${g_i} \in \mathcal{G}(A)$ such that ${x^*} + {g_i}$ is better than  ${x^*}$.

\item[(iv)] For any $g \in \mathcal{G}(A)$, an upper bound on the norm of Graver basis elements is given by
\begin{equation}\label{eq:graverupper}
{\left\| g \right\|_\infty } \leqslant \left( {n - r} \right)\Delta \left( A \right) \mbox{ and } \  {\left\| g \right\|_1 } \leqslant \left({n-r} \right)\left({r+1} \right)\Delta \left( A \right)
\end{equation}
where $r = rank(A) \leqslant m$  and $\Delta \left( A \right)$ is the maximum absolute value of the determinant of a square submatrix of $A$. 
\end{itemize}

%  Let us finish by pointing the interested reader to the appendix for a classical algorithm for computing Graver bases due to Pottier \cite{pottier_euclidean_1996} (historically, J. Graver did not provide an algorithm).
% \subsection{Graver basis, other definitions}
% In addition to the simple definition of Graver basis stated in Definition 2, reader can find two other equivalent definitions of what a Graver basis is, using Hilbert  and Groebner bases -- See in Appendix C. 
 
% Here, since mimicking the completion procedure using quantum annealer is impractical, we utilize the basic definition of Graver basis (Definition 2).  %I have an algo for this!  better remove it!
 
\subsection{Wide Applicability of Graver Bases as Optimality Certificates}\label{subsec:costcategories}
Beyond integer linear programs (ILP) with a fixed integer matrix, Graver bases as optimality certificates have now been generalized to include several nonlinear objective functions: 
\begin{itemize}
    \item Separable convex minimization \cite{murota_optimality_2004}: $\begin{array}{*{20}{c}}
  {\min }&{\sum\nolimits_i {{f_i}({c_i^Tx})} } 
\end{array}$ with $f_i$ is convex.

    \item Convex integer maximization (weighted) \cite{de_loera_convex_2009}: $\begin{array}{*{20}{c}}
  {\max }&{f({\mathbf{W}}x)}&{,{\mathbf{W}} \in {\mathbb{Z}^{d \times n}}} 
\end{array}$ with $f$ convex on $\mathbb{Z}^d$.
    
    \item  Norm $p$ (nearest to $x_0$) minimization  \cite{hemmecke_polynomial_2011}: $\begin{array}{*{20}{c}}
  {\min }&{{{\left\| {x - {x_0}} \right\|}_p}} 
\end{array}$. 
    
    \item  Quadratic minimization \cite{murota_optimality_2004, lee_quadratic_2012}: $\begin{array}{*{20}{c}}
  {\min }&{{x^T}Vx} 
\end{array}$ where $V$ lies in dual of quadratic Graver cone of $A$ 
    
    \item Polynomial minimization \cite{lee_quadratic_2012}: $\begin{array}{*{20}{c}}
  {\min }&{P(x)} \end{array}$ where  $P$ is a polynomial of degree $d$, that lies on cone $K_d(A)$, dual of $d^{th}$ degree Graver cone of $A$.
 
\end{itemize}
 It has been shown that only polynomially many augmentation steps are needed to solve such  minimization problems \cite{hemmecke_polynomial_2011} \cite{de_loera_convex_2009}. 
 Graver did not provide an algorithm for computing Graver bases; Pottier \cite{pottier_euclidean_1996} and Sturmfels \cite{sturmfels_variation_1997} provided such algorithms (see Appendix \ref{sec:classicalgraver}).  
 
 ~~\\
The use of test sets to study integer programs is a vibrant area of research. Another collection of test sets is based on Groebner bases. See \cite{conti_buchberger_1991,  tayur_algebraic_1995,bigatti_computing_1999, hosten_grin:_1995} and \cite{bertsimas_new_2000} for several different examples of the use of Groebner bases for integer programs.

\section{Graver Basis via Quantum Annealing} \label{sec:graverbasis}
Here we present our first hybrid quantum classical algorithm, for Graver basis,
 constructed from a subset of infinite set of kernels of matrix $A$ (obtained using the quantum annealer by solving a QUBO) and refining them classically:
\begin{itemize}
  \item \textbf{Quantum kernel:} Calculate a {\it subset} of the kernel as discussed below.
  \item \textbf{Classical filtering:} Refine the Graver basis using $\sqsubseteq$-minimal filtration on the kernel subset. 
\end{itemize}
 This two step procedure is repeated in a loop until all required Graver basis elements are acquired. In each iteration $i$ within this loop, we extract the Graver basis of the given integer matrix $A$ for a given bound $[L_g,U_g]$. These bounds are updated {\it adaptively} $[L_g^i,U_g^i]$ from one iteration to another.
% \subsection{Algorithms}

%\subsubsection{Hybrid QC Graver extraction algorithm}\label{subsec:algGraver}
~~\\
Each call of a hybrid quantum classical algorithm to extract Graver basis, inside such a bounded region, is as follows:
% For general unbounded Graver basis, the user only needs to chose large enough positive and negative upper and lower bounds.

~~\\
{\centering
\begin{minipage}{1.0\linewidth}
\begin{algorithm}[H]
\caption{Graver Extraction}\label{alg:GE}
\small{
\begin{algorithmic}[1]
\STATE {\bf inputs:} Integer matrix $A$, Graver basis lower ($L_{g}$) and upper ($U_{g}$) bound vectors
\STATE {\bf output:} Graver basis $\mathcal G(A)$, inside the bounded region $[L_{g},U_{g}]$
\STATE Initialize: Graver set $\mathcal G(A)$, Kernel set $K(A)$, Midpoint vector $M_0$, Encoding length vector $K_0$
\WHILE{$(iteration\ i)\ \Delta \left\| {G^{i}(A)} \right\| \leqslant \delta $}
\STATE Reshuffle columns of $A$, using random permutation $P_{i}$
\STATE Construct QUBO matrix $Q_{B}^{i}$, using equation (\ref{eq:ax0qubo}), (see \ref{subsec:ax0qubo})
\STATE Find optimal and suboptimal solutions $[x_i^o, x_i^{s}] = Quantum\ Solve(Q_B^{i})$,   (see \ref{sec:quantumannealer})
\STATE Revert variable orders (using $P_{i}$)
\STATE Post-process suboptimals $x_i^s$ into new optimals $x_i^{no}$, (see \ref{subsec:suboptimal})
\STATE Create unique kernel set $K_i(A) = x_i^o \bigcup x_i^{no}$
\STATE $\mathcal G(A) = \mathcal G(A)\bigcup {{K_i}} (A)$ and $K(A) = K(A)\bigcup {{K_i}} (A)$
\STATE \# Kernel to Graver, $\sqsubseteq minimal$ post-processing:
\FOR{$v_1,v_2$ {\bf in} $\mathcal G(A)$}
\STATE \bf{if} $v_{1} \sqsubseteq v_{2}$ \bf{then} $\mathcal G(A) = \mathcal G(A) \backslash v_2$
\ENDFOR
\STATE $\mathcal G(A) = Residual\ Procedure (\mathcal G(A), K_{i}(A))$; (see \ref{subsec:residual})
\STATE $[M_{i+1},K_{i+1}]= Adaptive\ Adjustments\ Procedure(M_i, K_i, x_i^s)$;   (see \ref{subsec:adaptive})
\ENDWHILE
\RETURN $\mathcal G(A)$
\end{algorithmic}
}
\end{algorithm}
\end{minipage}
}

\subsection{QUBOs for Kernels}\label{subsec:ax0qubo}
The mapping of the kernel problem into a QUBO is the essential quantum step in our hybrid approach. Each call of a quantum-annealing based QUBO solver can-- through multiple reads-- extract several (unique) degenerate ground states. We expect to be able to compute a large sampling of the  kernel of ${\bf{A}}$ in one or more calls (of 10000 reads each).

  ~~\\
%{\color{red} See my edits } \\
%To find the kernel of ${\bf{A}}$, 
We need to solve 
$\begin{array}{*{20}{c}}
  {{\mathbf{Ax}} = {\mathbf{0}},}&{{\mathbf{x}} \in {\mathbb{Z}^n}\begin{array}{*{20}{c}}
  ,&{{\mathbf{A}} \in {\mathbb{Z}^{m \times n}}}, 
\end{array}} 
\end{array}$ which is computationally hard, in a bounded region \cite{papadimitriou_complexity_1981}.
Instead, we look to find a large sampling of the non-zero elements of the kernel. To do this, we use quadratic unconstrained integer optimization (QUIO):
\begin{equation}\label{eq:intqubo}
\begin{array}{*{20}{c}}
  {\begin{array}{*{20}{c}}
  {\min }&{{{\mathbf{x}}^T}{{\mathbf{Q}}_{\mathbf{I}}}{\mathbf{x}}}&,&{{{\mathbf{Q}}_{\mathbf{I}}} = {{\mathbf{A}}^T}{\mathbf{A}}} 
\end{array}\begin{array}{*{20}{c}}
  ,&{{\mathbf{x}} \in {\mathbb{Z}^n}} 
\end{array}} \\ 
  {{{\mathbf{x}}^T} = \left[ {\begin{array}{*{20}{c}}
  {{x_1}}&{{x_2}}& \ldots &{{x_i}}& \ldots &{{x_n}} 
\end{array}} \right]\begin{array}{*{20}{c}}
  ,&{{x_i} \in \mathbb{Z}} 
\end{array}} 
\end{array}
\end{equation}

\subsubsection{Integer to binary encoding}\label{subsec:int2bin}
Because the quantum annealer solves quadratic unconstrained binary problems, we need to encode the integer variables into binary variables through an integer to binary mapping. Due to the limited number of qubits available, we need to be efficient in our encoding. We do so by adaptively allocating the qubits to the variables, so that the variables may have a different number allocated to them in any given iteration. In subsequent iterations, based on which variable appears to be in need, we readjust the allocation. We also re-center our variables adaptively to further increase the efficiency of encoding. We discuss this in a later subsection (see \ref{subsec:adaptive}).

~~\\
To map an integer variable to a binary, while keeping the problem quadratic, we use a linear mapping:
${x_i} = {\mathbf{e}}_i^T{X_i}$ where 
${\mathbf{e}}_i^T = \left[ {\begin{array}{*{20}{c}}
  {e_i^1}&{e_i^2}& \cdots &{e_i^{{k_i}}} 
\end{array}} \right] \in \mathbb{Z}_ + ^{{k_i}}$ 
is the linear encoding vector and 
$X_i^T = \left[ {\begin{array}{*{20}{c}}
  {{{X}_{i,1}}}&{{{X}_{i,2}}}& \cdots &{{{X}_{i,{k_i}}}}
\end{array}} \right] \in {\left\{ {0,1} \right\}^{{k_i}}}$
is the encoded binary vector that encodes integer value ${x_i}$.
The ${k_i}$ is the length (number of bits) of encoding vector ${{\mathbf{e}}_i}$, and depends on the linear encoding scheme that is in use.  For a specific integer length, the difference between the upper bound of ${x_i}$  to the lower bound of ${x_i}$ is ${\Delta _i} = U{x_i} - L{x_i}$; thus, the minimum ${k_i}$ is $\left\lceil {{{\log }_2}{\Delta _i}} \right\rceil $ for \textit{binary} encoding with the encoder vector
${\mathbf{e}}_i^T = \left[ {\begin{array}{*{20}{c}}
  {{2^0}}&{{2^1}}& \cdots &{{2^{{k_i}}}} 
\end{array}} \right]$, and the maximum ${k_i}$ is ${\Delta _i}$ for \textit{unary} encoding with the encoder vector 
${\mathbf{e}}_i^T = \overbrace {\left[ {\begin{array}{*{20}{c}}
  1&1& \cdots &1 
\end{array}} \right]}^{{k_i}}$.
Therefore,
\begin{equation}\label{eq:encoding}
{\mathbf{x}} = {\mathbf{L}} + {\mathbf{EX}} = \left[ {\begin{array}{*{20}{c}}
  {L{x_1}} \\ 
  {L{x_2}} \\ 
   \vdots  \\ 
  {L{x_n}} 
\end{array}} \right] + \left[ {\begin{array}{*{20}{c}}
  {{\mathbf{e}}_1^T}&{{{\mathbf{0}}^T}}& \cdots &{{{\mathbf{0}}^T}} \\ 
  {{{\mathbf{0}}^T}}&{{\mathbf{e}}_2^T}& \cdots &{{{\mathbf{0}}^T}} \\ 
   \vdots & \vdots & \ddots & \vdots  \\ 
  {{{\mathbf{0}}^T}}&{{{\mathbf{0}}^T}}& \cdots &{{\mathbf{e}}_n^T} 
\end{array}} \right]\left[ {\begin{array}{*{20}{c}}
  {{X_1}} \\ 
  {{X_2}} \\ 
   \vdots  \\ 
  {{X_n}} 
\end{array}} \right]
\end{equation}
where ${\mathbf{E}}$ is a $(n \times \sum\limits_{i = 1}^n {{k_i}} )$  matrix that encodes the integer vector ${\mathbf{x}}$ into binary vector ${\mathbf{X}} \in {\left\{ {0,1} \right\}^{\sum {{k_i}} }}$  and ${\mathbf{L}}$ is a $(n \times 1)$ vector containing the integer variable’s lower bounds. (If one chooses the same encoding size for all integer variables
$\begin{array}{*{20}{c}}
  {{\mathbf{e}} = {{\mathbf{e}}_i}}&,&{k = {k_i}}&,&{\forall i = 1, \ldots ,n},  
\end{array}$
 then the $(n \times nk)$ encoding matrix simplifies to
${\mathbf{E}} = {\mathbf{e}} \otimes {{\mathbf{I}}_n}$
, where $ \otimes $ is the Kronecker product and ${{\mathbf{I}}_n}$ is the identity matrix of size $n$.) Inserting (\ref{eq:encoding}) into (\ref{eq:intqubo}):\\
\begin{equation}\label{eq:quboencoded}
{{\mathbf{x}}^T}{{\mathbf{Q}}_{\mathbf{I}}}{\mathbf{x}} = {\left( {{\mathbf{L}} + {\mathbf{EX}}} \right)^T}{{\mathbf{Q}}_{\mathbf{I}}}\left( {{\mathbf{L}} + {\mathbf{EX}}} \right) = {{\mathbf{X}}^T}{{\mathbf{E}}^T}{{\mathbf{Q}}_{\mathbf{I}}}{\mathbf{EX}} + \left( {{{\mathbf{X}}^T}{{\mathbf{E}}^T}{{\mathbf{Q}}_{\mathbf{I}}}{\mathbf{L}} + {{\mathbf{L}}^T}{{\mathbf{Q}}_{\mathbf{I}}}{\mathbf{EX}}} \right) + {{\mathbf{L}}^T}{{\mathbf{Q}}_{\mathbf{I}}}{\mathbf{L}}
\end{equation}
The last term in equation (\ref{eq:quboencoded}) is constant. Because $x_i^2 = x_i$ for binary variables, the middle linear terms can be rewritten as  ${{\mathbf{X}}^T}diag\left( {2{{\mathbf{L}}^T}{{\mathbf{Q}}_{\mathbf{I}}}{\mathbf{E}}} \right){\mathbf{X}}.$
Therefore the quadratic unconstrained binary optimization formulation simplifies to, 
\begin{equation}\label{eq:ax0qubo}
\begin{array}{*{20}{c}}
  {\begin{array}{*{20}{c}}
  {\min }&{{{\mathbf{X}}^T}} 
\end{array}{{\mathbf{Q}}_{\mathbf{B}}}{\mathbf{X}}\begin{array}{*{20}{c}}
  ,&{{{\mathbf{Q}}_{\mathbf{B}}} = {{\mathbf{E}}^T}{{\mathbf{Q}}_{\mathbf{I}}}{\mathbf{E}} + diag\left( {2{{\mathbf{L}}^T}{{\mathbf{Q}}_{\mathbf{I}}}{\mathbf{E}}} \right)} 
\end{array}} \\ 
  {{\mathbf{X}} \in {{\left\{ {0,1} \right\}}^{nk}}\begin{array}{*{20}{c}}
  ,&{{{\mathbf{Q}}_{\mathbf{I}}} = {{\mathbf{A}}^T}{\mathbf{A}}} 
\end{array}} 
\end{array}
\end{equation}

\subsubsection{Sampling the kernel}\label{subsec:sampling}

The QUBO in equation (\ref{eq:ax0qubo}) has inherent symmetry in  ${{\mathbf{Q}}_{\mathbf{I}}}$ for each of the variables. For a randomly chosen integer matrix ${\mathbf{A}}$, the valleys for all degenerate solutions should have similar spreads. This gives us optimism that with certain variable perturbations and some post-processing of near-optimal solutions, we should be able to sample well. Additionally, because we solve (\ref{eq:ax0qubo}) in a sequence of adaptive D-Wave calls, fine-tuning the bit sizes for each variable and  targeting our kernel search, we expect that although the D-Wave annealer is not an uniform sampler (in the theoretical sense, see \cite{mandra_exponentially_2017}), we should nevertheless be quite successful. 

  ~~\\
We should mention that applying any modification that changes ${{\mathbf{Q}}_{\mathbf{B}}}$ requires a new embedding of the modified problem graph into the sparse hardware graph of the D-Wave processor, which is a downside. Using  D-Wave SAPI in-house embedder program, the embedder starts from scratch each time, which creates an unknown delay in each iteration of the main loop. We have designed a new approach that transforms the embedding of any problem graph into any hardware graph into an equational system of quadratic integer equations and solves them by powerful tools of algebraic geometry and Groebner bases
\cite{cmuag}. This approach has the property that minor changes in ${{\mathbf{Q}}_{\mathbf{B}}}$  result in only minor change in equations, thus the Groebner bases solver still utilizes the majority of simplified equations from the previous pool of equations that leads to shorter embedding time for minor changes in ${{\mathbf{Q}}_{\mathbf{B}}}$. We have not used this new embedding approach in this paper, however, as our goal was to see what the available hardware and SAPI could do.

\subsection{Post-Processing}\label{subsec:postprocessing}
We have developed two post-processing steps to be done classically.
\subsubsection{Systematic post-processing of near-optimal solutions into optimal}
Implementation of the model in equation (7) on the D-Wave quantum annealing solver (DW2000Q, solver id: C16-VFYC) shows promising results in extracting degenerate kernel solutions, especially when using minimize energy chainbreak strategy (see Appendix \ref{sec:quantumannealer}). For a random selection of integer matrices  with various sizes, one run of the processor with approximately 10000 reads (annealing time 290 {\it msec})  returns $\approx 100 - 150$ unique (non-similar) optimal and suboptimal solutions. Among these, $\approx 10 - 20$ are optimal degenerate solutions and the rest are suboptimal solutions.

  ~~\\
Analysis of these suboptimal solutions shows  that, remarkably, the majority (over $87\%$) of them have very small errors, and so are near-optimal. This motivated us to design a post-processing procedure to extract more degenerate solutions from the existing suboptimal solutions.  Because this post-processing is based on our known problem ($Ax=0$), the operations required to reach the optimal solutions are systematic and with known predictable polynomial overhead. 

\subsubsection{Distribution of optimal and near-optimal solutions: Experimental results}
\label{sec:suboptimal}
% As said many of the solutions acquired from D-Wave processor are sub-optimal. We find out that, while using the default \textit{minimize energy} chainbreak strategy, majority of them have energies which are very close to the optimal energy.
We chose a selection of random integer matrices ${\mathbf{A}} \in {\mathbb{Z}^{m \times n}}\begin{array}{*{20}{c}}
  ,&{ - 10 \leqslant {a_{ij}} \leqslant 10} \end{array}$; 
  ten matrices from each size 
 $\left( {m \times n} \right) \in \left\{ {\left( {2 \times 5} \right),{\text{ }}\left( {3 \times 5} \right),{\text{ }}\left( {3 \times 6} \right),{\text{ }}\left( {3 \times 7} \right),{\text{ }}\left( {4 \times 8} \right),{\text{ }}\left( {5 \times 8} \right)} \right\}$.
 Having a fixed embedding for each size category, we acquired the unique solutions of each call ($10000$ reads per call, annealing time of $290nsec$), and evaluated the percentage of unique solutions, which fall into any of the seven solution categories: zero error (optimal solutions), 1 to 5 total sum errors, and greater than or equal to $6$ total sum errors.
 
\begin{table}[H]
\centering
\begin{tabular}{|l|l|l|l|l|l|l|l|}
\hline
$\ddots$ & $(2\times5)$ & $(3\times5)$ & $(3\times6)$ & $(3\times7)$ & $(4\times8)$ & $(5\times8)$ & Overall\\
\hline
0 & 15 & 12 & 10 & 13 & 12 & 9 & 11.8 \\
\hline
1 & 24 & 28 & 26 & 31 & 30 & 33	& 28.7 \\
\hline
2 & 31 & 28 & 27 & 24 & 21 & 25 & 26.0 \\
\hline
3 & 20 & 21 & 25 & 19 & 20 & 17 & 20.3 \\
\hline
4 & 7 & 8 & 10 & 9 & 9 & 8 & 8.5 \\
\hline
5 & 2 & 1 & 0 & 3 & 4 & 5 & 2.5 \\
\hline
$\geqslant$6 & 1 & 2 & 2 & 1 & 4 & 3 & 2.2 \\
\hline
\end{tabular}
\caption{\label{tab:subopt} Optimal and suboptimal solution percentages for various size matrices.}
\end{table}

~~\\
As seen in   Table \ref{tab:subopt}, over $75\%$ of unique solutions, fall into the categories with total sum errors 1 to 3. Overall, $87\%$ of unique solutions  are either optimal solutions or near-optimal.

\subsubsection{Combining near-optimal solutions}\label{subsec:suboptimal}
Assume ${{\mathbf{x}}_u} \in {\mathbb{Z}^{n \times N}}$ is the unique solution matrix (each column is a unique read), acquired by filtering out the repeated solutions out of all reads from the quantum annealer.

~~\\
The matrix ${\mathbf{A}}{{\mathbf{x}}_u} = {E_r}$ contains the $b$  values in each of $N$ columns. We expect that $b = {\bf{0}}$ for optimal solutions, and $b \ne {\bf{0}}$ but ${{\mathbf{1}}_n}\left| b \right|$  small for nonoptimal solutions.  We sorted and separated the columns of ${{\mathbf{x}}_u}$ based on their corresponding sum of absolute value error in ${E_r}$. Because the majority of reads have no maximum error greater than 3, we categorized solutions into four groups of  0, 1, 2, and 3 total sum errors. The solutions with a total sum of 0 obviously do not need post-processing.
\paragraph{Post-processing for the solutions with a total sum error = 1}
For each row $
  {{r_i}},\, {1\leq i \leq m} 
$ all  columns with +1 error and all columns with -1 error consist two sub-groups. For each row, thus, we have three sets of operations: Subtracting all +1 columns pairwise, subtracting all -1 columns pairwise, and adding any of columns from +1 to any of columns from -1 pairwise.
The above-mentioned operations, on average for all $m$ rows, cost $O\left( {\frac{{3N_{u1}^2}}{{4m}}} \right)$ pairwise column additions, where ${N_{u1}}$ is the number of unique solutions with a total sum error of 1.

\paragraph{Post-processing for the solutions with total sum error = 2}
The absolute sum error =2 situation has two cases. The first one is similar to the previous case; for each of the $m$ rows, we have either +2 or -2  (and of course 0), which we will resolve similarly. In the second case, for each pair of rows, $\begin{array}{*{20}{c}}
  {{r_i},{r_j}}&,&{i,j = 1, \ldots ,m} 
\end{array}$, among $m(m - 1)/2$ possible pairs, we have four sub-groups of errors $({r_i},{r_j})$ that are (+1, +1), (+1, -1), (-1, +1), and (-1, -1). For each sub-group any pairwise subtraction results to zero. Additionally, we have two sub-groups of {(+1, +1), (-1, -1)} and {(+1, -1), (-1, +1)} that can have pairwise subtraction to result in zero.
The above-mentioned operations, for all $m$ rows, on average roughly cost $O\left( {\frac{{3N_{u2}^2}}{{4m}}} \right)$ column pairwise additions to resolve +2, -2 errors and $O\left( {\frac{3}{8}N_{u2}^2} \right)$ pairwise additions to resolve the eight remaining categories of various +1, -1 errors, where ${N_{u2}}$ is the number of unique solutions with a total sum error of 2.

\paragraph{Post-processing for the solutions with total sum error = 3}
This situation has three general categories of errors.
As before, if for each of m rows, in each column we have either +3 or -3, this case is similar to the first case (and the first case in the second situation) and is resolved similarly.
The second case here has similarities with the second category of the second case. It consists of 4 subgroups on $m(m-1)/2$ pairs of rows, that have (-2, -1), (-2, +1),  (+2, -1), (+2, +1). Likewise, it needs in-group subtraction for each of the four groups and two pairwise additions between the subgroups {(-2, -1), (+2, +1)} and {(-2, +1), (+2, -1)}.
The third category consists of columns, which have one of the eight groups of (-1, -1, -1), (-1, -1, +1), (-1, +1, -1), (-1, +1, +1), (+1, -1, -1), (+1, -1, +1), (+1, +1, -1), (+1, +1, +1). For every triple row among $\left( {\begin{array}{*{20}{c}}
  m \\ 
  3 
\end{array}} \right)$ 
possible row combinations, we have eight in-group pairwise column subtractions, plus four out-group pairwise additions between {(-1, -1, -1), (+1, +1, +1)}, {(-1, -1, +1), (+1, +1, -1)}, {(-1, +1, -1), (+1, -1, +1)}, {(-1, +1, +1), (+1, -1, -1)}.
These operations, for all $m$ rows, on average roughly cost $O\left( {\frac{{3N_{u3}^2}}{{4m}}} \right)$ column pairwise additions to resolve +3, -3 errors and $O\left( {\frac{3}{8}N_{u3}^2} \right)$  pairwise additions to resolve the eight category of +/-2, +/-1 errors, and maximum $O\left( {\frac{2}{9}mN_{u3}^2} \right)$  for resolving 12 cases with three +/-1 error in a column, which ${N_{u3}}$  is the number of unique solutions with total sum error equal to 3.

\paragraph{Notes:} 

~~\\
(a) The routine used in case sum error =1  is reused  (with different values) in the first cases of sum error=2 and sum error=3. Similarly, the routine used in the second category of sum error =2 is used in the second case of sum error =3.

~~\\
(b) This systematic post-processing procedure can go to higher levels of errors, but detailed analysis of all unique solutions using the D-Wave processor shows that the majority of sub-optimal solutions have sum errors up to 3 (with low chainbreak rate, also while using the minimize energy chain break strategy), so we did not implement the post-processing for the higher error values.

~~\\
(c) Notice that the span of values for post-processed solutions becomes twice the original span of values set by encoding size, because there are additions of columns, thus it acts like adding one extra bit to each variable.

\subsection{Using Residual Kernel to Obtain More Graver Bases Elements}\label{subsec:residual}
After acquiring the kernel set $K_i$ in call iteration $i$, by sifting the $\sqsubseteq$ minimal elements of the kernel into Graver bases elements $G_i$, many of the kernel elements that are not part of the Graver bases (or any multiple of them) are left unused ($K_i^*: K_i = G_i \cup K_i^*$). They are either:  (1) a linear combination of the Graver bases elements already found,  or (2) a linear combination of found and not yet found Graver bases elements. The second case is of interest. Inspired by Pottier's algorithm, the post-processing procedure to acquire the not yet found Graver bases elements, from these extra kernel elements, is as follows:

  ~~\\
{\centering
\begin{minipage}{1.0\linewidth}
\begin{algorithm}[H]
\caption{Residual Kernel}\label{alg:PROC1}
\small{
\begin{algorithmic}[1]
\STATE \textbf{input} (iteration $i$) Partial Graver basis set $\mathcal G_i$ and existing kernel set $K_i^*$
\STATE \textbf{output} Graver basis set $\mathcal G_i^r \subseteq \mathcal{L}\backslash \left\{ {\mathbf{0}} \right\}$
\STATE Initialize symmetric set: $K\leftarrow K_i^* \cup ( - K_i^*)$
\STATE Generate (in orthant $\theta _j$) the vector set: $C_j \leftarrow \bigcup\limits_{f^j \in K , g^j \in \mathcal  G_i} {\left\{ {f^j + g^j} \right\}}$
\STATE Collecting residues from all existing orthants: ${{\mathbf{C}}_i}: = \bigcup\limits_j {{C_j}}$
\STATE Extracting the $\sqsubseteq$ conformal minimal elements: ${\mathcal G_i^r} =  \sqsubseteq \left( {{\mathcal G_i} \cup {{\mathbf{C}}_i}} \right)$ 
\RETURN $\mathcal G_i^r$
\end{algorithmic}
}
\end{algorithm}
\end{minipage}
}
  ~~\\
This simple post-processing procedure is implemented inside the Graver extraction algorithm.

\subsection{Adaptive Centering and Encoding}\label{subsec:adaptive}
% \subsection{Adaptive adjustment of encoding size and center}
 We formulated instances of the problem QUBO in such a way that one can change each integer variable’s number of encoding bits ${k_i}$ allocated to variable ${x_i}$. We also formulated the QUBO such that the encoding's base value, either the lower point ${L_{{x_i}}}$ or the  central point (see equations \ref{eq:encoding}  to \ref{eq:ax0qubo}) can be adjusted.

~~\\
In each iteration, we check the previous iteration solution's possible border (corner) values among suboptimal unique solutions. A solution has a border value if its value is either the upper or the lower value of the encoding range. For example, if a variable has 4 bits encoding, the solution range (considering a zero encoding center) is between $-8$ to  $+7$; thus, if the suboptimal solution is either $-8$ or $+7$, it is on the border. If a solution in iteration $i$ is on the border, we expect that a better  solution lies beyond this border.  So, we can adapt either (or both) the center and size of this variable in the next iteration. In the previous example having a value of $+7$ is an indication that in iteration $i + 1$, the center could be moved toward $+7$. Instead of just moving the center, we can also increase the encoding length of that variable by one, and decrease the encoding length of another variable (which has small enough values).

~~\\
We initially use the same encoding length and zero center for all variables. To prevent sudden changes in encoding length and base, we employed a moving average-based filtering method. 
A simplified version of the adaptive adjustment is as follows: 
  ~~\\
{\centering
\begin{minipage}{1.0\linewidth}
\begin{algorithm}[H]
\caption{Adaptive Adjustments}\label{alg:PROC2}
\small{
\begin{algorithmic}[1]
\STATE \textbf{input} (iteration $i$) Middle points vector $M_{i}$, Encoding lengths vector $K_{i}$, suboptimals vector $x_i$
\STATE \textbf{output} Middle points vector $M_{i+1}$, Encoding lengths vector $K_{i+1}$
\STATE Calculate saturated borders: $right\ border$ and $left\ border$ based on $M_i$ and $K_i$
\STATE \# Adaptive adjustment of middle points
\IF {$(x_i == right\  border(M_{i}, K_{i}))$} 
\STATE $M_{i+1} = M_{i} +1$
\ELSIF {$(x_i == left\ border(M_{i}, K_{i}))$}
\STATE $M_{i+1} = M_{i} -1$
\ENDIF
\STATE \# Adaptive adjustment of encoding lengths
\IF {$\left| {x_i - M_i} \right| \leqslant {2^{K_i - 1}}$}
\STATE $K_{i+1} = K_i -1$
\ENDIF 
\RETURN $M_{i+1}$ and $K_{i+1}$
\end{algorithmic}
}
\end{algorithm}
\end{minipage}
}

\subsection{Numerical Results}\label{subsec:gravertests}

We provide several examples illustrating the calculation of Graver bases using the D-Wave quantum annealer. We used the chain break post-processing {\it minimize energy} first, and then redid the calculations using {\it majority vote} (see \ref{subsec:chainbreak} for definitions).
\subsubsection{Example 1: Four coin problem}\label{example1}
%  Topical subheadings are allowed.
Consider the problem of four coins discussed in \cite{sturmfels_algebraic_2003}, which asks for replacing a specific amount of money with  the correct combination of pennies (1 cent), nickels (5 cents), dimes (10 cents), and quarters (25 cents) and a fixed total  sum of coins.

\begin{center}
$A = \left[ {\begin{array}{*{20}{c}}
1&1&1&1\\
1&5&{10}&{25}
\end{array}} \right]$
\end{center}
Allocating $k=4$ bits for each of the variables, hence limiting the integer variable span to $[-8, +7]$ using binary encoding, one call obtained four out of five Graver basis without any post-processing. Obviously, because one of the basis elements is out of encoding range, it could not be found. However, because we post-process near optimal solutions, using only $sum error =1$, post-processing extracted the other basis element. The Graver basis, also used in our illustrative example (\ref{subsec:illustrative}), is
\begin{center}
$\mathcal G(A)= \left[ {\begin{array}{*{20}{c}}
0&5&5&5&5\\
3&{ - 9}&{ - 6}&{ - 3}&0\\
{ - 4}&4&0&{ - 4}&{ - 8}\\
1&0&1&2&3
\end{array}} \right]$
\end{center}
Changing the chain break to majority vote, we still extract all the Graver basis using two D-Wave calls (average $52\%$ chainbreak).

\subsubsection{Example 2: Variation problem}
The following $(3 \times 6)$ matrix in an example given in \cite{sturmfels_variation_1997}:

$$A = \left[ {\begin{array}{*{20}{c}}
  2&1&1&0&0&0 \\ 
  0&1&0&2&1&0 \\ 
  0&0&1&0&1&2 
\end{array}} \right]$$
Here $n = 6$, $r = 3$ and $\Delta(A) = 2$. Based on Proposition 1-iv (equation \ref{eq:graverupper}), ${\left\| g \right\|_\infty } \leqslant 6 $, we start by allocating $k=3$ bits for each of the six variables, hence limiting the integer variable span to $[-4, +3]$ using binary encoding. One D-Wave call extracted $10$ Graver basis elements. Increasing $k$ to 4, still gives us the same $10$ Graver basis elements. 

$$
\mathcal G(A) =  
\begin{bmatrix}
 { 0}  &   0  &   0  &   0  &   1  &   1  &   1  &   1  &   1   &  1\\
  0   &  1  &   1  &   2  &   - 2   &  - 2   &  - 1  &   - 1    & 0  &   0\\
  0   &  - 1   &  - 1  &   - 2  &   0  &   0   &  - 1  &   - 1  &   - 2   &  - 2\\
  1   &  - 1  &   0   &  - 1    & 0   &  1   &  0  &   1  &   - 1  &   0\\
  - 2  &   1   &  - 1  &   0    & 2  &   0  &   1    & - 1  &   2  &   0\\
  1  &   0   &  1  &   1   &  - 1 &    0  &   0  &   1  &   0   &  1
\end{bmatrix}$$
Changing the chain break to majority vote, we extract all the Graver basis in one D-Wave calls (average only $8\%$ chainbreak, due to the small size of the problem, thus small length of chains).

\subsubsection{Example 3: Snake-wise prime number}
Consider the problem of filling a matrix snake-wise with prime numbers from  Ref. \cite{sturmfels_algebraic_2003}, but with smaller matrix size of $(2 \times 5)$:

\begin{center}
$A = \left[ {\begin{array}{*{20}{c}}
1&2&3&5\\
{15}&{13}&{11}&7
\end{array}} \right]$
\end{center}
  ~~\\
This matrix has 22 Graver basis elements that were extracted in 3 calls using 6 bits of binary encoding and utilizing all three post-processings.
\begin{gather*}
\mathcal G(A) =  \left[ {\begin{array}{*{22}{c}}
   0 & {17} & {17} & {17} & {17} & {17} & {17} & {17} & {17} & {17} & {17} & {17} & {17} & {17} & {17} & {17} & {17} & {17} & {17} & {17} & {17} & {17} \\
   2 & { - 40} & { - 38} & { - 36} & { - 34} & { - 32} & { - 30} & { - 28} & { - 26} & { - 24} & { - 22} & { - 20} & { - 18} & { - 16} & { - 14} & { - 12} & { - 10} & { - 8} & { - 6} & { - 4} & { - 2} & 0  \\ 
   { - 3} & {21} & {18} & {15} & {12} & 9 & 6 & 3 & 0 & { - 3} & { - 6} & { - 9} & { - 12} & { - 15} & { - 18} & { - 21} & { - 24} & { - 27} & { - 30} & { - 33} & { - 36} & { - 39}  \\
   1 & 0 & 1 & 2 & 3 & 4 & 5 & 6 & 7 & 8 & 9 & {10} & {11} & {12} & {13} & {14} & {15} & {16} & {17} & {19} & {19} & {20}  \\
\end{array}} \right]
\end{gather*}

  ~~\\
Choosing the higher sizes for the snake-wise prime number matrix increases the Graver basis element range, which needs a higher number of binary encoding bits, thus makes the problem unembeddable in the current processor.
Changing the chain break to majority vote, we could extract all the  Graver basis in 8 D-Wave calls (average $61\%$ chainbreak).

\subsubsection{Example 4: Some random matrices of size (2 $\times$ 5)}
The following matrix from Ref. \cite{onn_nonlinear_2010}:
\begin{center}
$A = \left[ {\begin{array}{*{20}{c}}
{ - 8}&{ - 9}&1&0&4\\
4&1&8&{ - 2}&5
\end{array}} \right]$
\end{center}
This matrix has 107 Graver basis elements. Using 6 binary encoding followed by three level post-processing and adaptive encoding of the center, in five D-Wave calls, 98 of the Graver basis were acquired. 
 
  ~~\\
Choosing a random integer matrix of the same size, with integer values spanning between -10 and 10, roughly 90 percent of the Graver basis elements were acquired in less than 10 D-Wave calls.
Changing the chain break to majority vote, we could extract only $12\%$ of the Graver basis in 10 D-Wave calls (average $72\%$ chainbreak).

\subsubsection{Example 5: Four cycle binary model}

The four-cycle binary model matrix size (16 $\times$ 16) in \cite{sturmfels_algebraic_2003}:

\begin{center}
$A = \left[ {\begin{array}{*{20}{c}}
1&1&1&1&0&0&0&0&0&0&0&0&0&0&0&0\\
0&0&0&0&1&1&1&1&0&0&0&0&0&0&0&0\\
0&0&0&0&0&0&0&0&1&1&1&1&0&0&0&0\\
0&0&0&0&0&0&0&0&0&0&0&0&1&1&1&1\\
1&1&0&0&0&0&0&0&1&1&0&0&0&0&0&0\\
0&0&1&1&0&0&0&0&0&0&1&1&0&0&0&0\\
0&0&0&0&1&1&0&0&0&0&0&0&1&1&0&0\\
0&0&0&0&0&0&1&1&0&0&0&0&0&0&1&1\\
1&0&0&0&1&0&0&0&1&0&0&0&1&0&0&0\\
0&1&0&0&0&1&0&0&0&1&0&0&0&1&0&0\\
0&0&1&0&0&0&1&0&0&0&1&0&0&0&1&0\\
0&0&0&1&0&0&0&1&0&0&0&1&0&0&0&1\\
1&0&1&0&1&0&1&0&1&0&1&0&1&0&1&0\\
0&1&0&1&0&1&0&1&0&0&0&0&0&0&0&0\\
0&0&0&0&0&0&0&0&1&0&1&0&1&0&1&0\\
0&0&0&0&0&0&0&0&0&1&0&1&0&1&0&1
\end{array}} \right]$
\end{center}
  ~~\\
The rank of this matrix is 9, and the maximum of all sub-determinants is 1, thus based on the upper bound evaluation, Proposition 1-iv, equation (\ref{eq:graverupper}) in section (\ref{sec:background}), the Graver basis has no element greater than 7. It  has 106 elements, and all of them were acquired with three D-Wave calls by setting annealing time to the maximum value of  $280 \mu sec $ (while having 10000 reads).  Interestingly, only one D-Wave call was needed when using the annealing time of $1 \mu sec$.

  ~~\\
By creating and storing the fixed size embedding of a complete graphs of size 64, we can eliminate the embedding time. This ultimately reduces the total computation time of Graver basis of this matrix with current D-Wave processors to only $4.38sec$. The majority of this time is due to the quantum processing unit {\it access} time. 
% If one can run the entire algorithm using a classical processor connected to quantum processor, the overall time will reduce substantially, especially in bigger problems.
 The total time derived by the classical program 4ti2 for this problem is about 3.0 seconds. This gives us optimism that first generation quantum annealers (with a first attempt at hybrid algorithms) are not that far behind best-in-class classical algorithms.
 Changing the chain break to majority vote, we could extract only $9\%$ of the Graver basis in 10 D-Wave calls (annealing time $280 \mu sec$) with the average $80\%$ chainbreak. 

\subsubsection{Example 6: Random low span integer matrices of size (4$\times$8)}
Using random integer matrices of size (4$\times$8), with a low integer value span of (-4 , +4), and using 6 binary encoding, we could calculate almost all of the Graver basis elements in less than 5 D-Wave calls ($10000$ reads, annealing time $280 \mu sec$). Reducing the annealing time to a minimum of $(1-10) \mu sec$ increased the number of unique solutions per call, but with the price of increased percentage of chainbreaks.
Changing the chain break strategy to majority vote, and returning the annealing time back to $280 \mu sec$ and specific embeddings, we could extract $15\%$ of the Graver basis with 10 D-Wave calls, while the average chainbreak was $65\%$. 
% New test should be repeated for vote

\subsection{Discussion}\label{subsec:graverdiscussion}
In general, we did not expect the current version of the quantum annealer (D-Wave 2000Q) to be competitive against classical computing that has decades of innovations and implementation improvements. Our motivation in this work was to demonstrate the possibility of calculating the Graver bases and subsequently optimization of nonlinear convex costs, using a quantum-classical approach.
\subsubsection{Chainbreak}
As we observe from the examples, the percentage of Graver basis acquired using D-Wave processor is mainly dependent on chainbreak strategy (see \ref{subsec:chainbreak}). The reality is that with the current sparse Chimera structure, the majority of problems embedded end up becoming very dense, and thus have long chains, which are prone to be broken. For such problems when using \textit{minimize energy}, we obtained many unique solutions (1000+), as many as just one order of magnitude less than number of reads (10000), which makes our hybrid approach competitive with classical methods. While using \textit{majority vote}, however, the number of unique solutions drops significantly, sometimes by two or three orders of magnitude from 10000, making it difficult be competitive to classical algorithms.

\subsubsection{Communication delays}
The communication delay from our local computer to the D-Wave machine, in addition the unknown job queuing delay before quantum processing, is an unknown idle time. This happens between classical and quantum computations in each iteration.  These prevent us from carrying out an exact time comparison between our hybrid algorithm and classical Graver basis software packages, such as 4ti2 \cite{noauthor_4ti2--software_nodate}. The lower bound for the computation time is the quantum processor total annealing time, which is low, and the upper bound is the overall time, which contains all extra idle times. One way of keeping score is to track the number of quantum calls, each with a maximum of under 3 seconds (10000 reads with a maximum anneal time of 290 $\mu sec.$)

  ~~\\
Detailed tests of the 4ti2 package (and other classical solvers) shows that when the problem size approaches 120 variables (for random binary matrices), the classical method takes a long time to find the Graver basis elements. In other words, it reaches its {\it exponential crunch point}. We believe that for such problems, the majority of Graver basis elements can be found with 4-5 bits of encoding for each variable. Thus, if we had a quantum annealer with a higher graph density, or higher number of qubits and the same density, which could embed problems of size 200, we could surpass the current classical approaches. (We also see this in our next section with respect to optimization.)  This is not unrealistic with the next generation of D-wave processors being developed \cite{boothby_next-generation_2018}.

\section{Hybrid Quantum Classical Algorithm for Non-Linear Integer Optimization}
\label{sec:nonlinearopt}

\subsection{Algorithm Structure}\label{subsec:algnonlin}
 For solving problem types given in equation (2), we do not necessarily need all Graver basis elements at the beginning of the algorithm. Therefore, we propose the following:
 \begin{itemize}
  \item \textbf{Truncated Graver basis:} Calculate the Graver basis, just for the bound set by difference of the upper and lower bound.
  
 \item \textbf{Initial feasible point(s):} Using a call to  the quantum annealer, find the set of feasible solutions and select one with minimum cost, or many, to increase redundancy.

 \item \textbf{Augmentation:} Successively augmenting to better solutions, as discussed in Proposition 1-(iii).
\end{itemize}
% \subsubsection{Hybrid QC Graver optimization algorithm}
Our hybrid quantum classical algorithm to find the optimum value of a nonlinear integer program is as follows (simplified). This algorithm uses Algorithm \ref{alg:GE} to extract Graver basis.

{\centering
\begin{minipage}{1.0\linewidth}
\begin{algorithm}[H]
\caption{Graver Optimization}
\label{alg:GOA}
\begin{algorithmic}[1]
\STATE {\bf inputs:} Matrix $A$, vector $b$, cost function $f(x)$, bounds $[l,u]$, as in Equation (\ref{eq:gennonlin})
\STATE {\bf output:} Augmentation evaluation of the objective function $f(x)$ and reaching to global solution $x^{*}$.
\STATE Using Algorithm \ref{alg:GE} inputs: $A$, $L_g=-(u-l)$, $U_g=+(u-l)$, extract truncated $\mathcal G(A)$
\STATE Finding a set of feasible solutions using Equation (\ref{eq:axbqubo}), while maintaining ${l \leqslant x \leqslant u}$ using middle point vector $M$ and encoding length vector $K$.
\STATE Sort the feasible solutions from \textit{low} to \textit{high} according to evaluated objective functions.
\STATE Select the feasible solution $x_{0}$ corresponding to the \textit{low} objective value ($x = x_{0}$).
\WHILE{$g \in \mathcal G(A)$}
\IF{$l \leqslant(x+g)\leqslant u$ and $f(x+g) < f(x)$}
\STATE $x=x+g$
\ENDIF
\ENDWHILE
\RETURN $x^{*}=x$
\end{algorithmic}
\end{algorithm}
\end{minipage}
}
\\

\subsection{QUBO for Obtaining Feasible Solutions}\label{subsec:axbqubo}
We first obtain a feasible solution for ${\mathbf{Ax}} = {\mathbf{b}}$ as follows. Note:
~~\\
\begin{equation}
{\left( {{\mathbf{Ax}} - {\mathbf{b}}} \right)^T}\left( {{\mathbf{Ax}} - {\mathbf{b}}} \right) = {{\mathbf{x}}^T}{{\mathbf{A}}^T}{\mathbf{Ax}} - \left( {{{\mathbf{x}}^T}{{\mathbf{A}}^T}{\mathbf{b}} + {{\mathbf{b}}^T}{\mathbf{Ax}}} \right) + {{\mathbf{b}}^T}{\mathbf{b}}
\end{equation}
The last term is constant, and the two middle terms are equal; hence to solve ${\mathbf{Ax}} = {\mathbf{b}}$ one needs to solve the quadratic unconstrained integer optimization (QUIO):
~~\\
\begin{equation}
\begin{array}{*{20}{c}}
  {\begin{array}{*{20}{c}}
  {\min }&{{{\mathbf{x}}^T}{{\mathbf{Q}}_{\mathbf{I}}}{\mathbf{x}} - 2{{\mathbf{b}}^T}{\mathbf{Ax}}}&,&{{{\mathbf{Q}}_{\mathbf{I}}} = {{\mathbf{A}}^T}{\mathbf{A}}} 
\end{array}} \\ 
  {{\mathbf{x}} \in {\mathbb{Z}^n}} 
\end{array}
\end{equation}
After applying the integer to the binary encoding of equation (\ref{eq:encoding}), the quadratic unconstrained binary optimization (QUBO) simplifies to.\\
\begin{equation}\label{eq:axbqubo}
\begin{array}{*{20}{c}}
  {\begin{array}{*{20}{c}}
  {\min }&{{{\mathbf{X}}^T}} 
\end{array}{{\mathbf{Q}}_{\mathbf{B}}}{\mathbf{X}}\begin{array}{*{20}{c}}
  ,&{{{\mathbf{Q}}_{\mathbf{B}}} = {{\mathbf{E}}^T}{{\mathbf{Q}}_{\mathbf{I}}}{\mathbf{E}} + 2diag\left[ {\left( {{{\mathbf{L}}^T}{{\mathbf{Q}}_{\mathbf{I}}} - {{\mathbf{b}}^T}{\mathbf{A}}} \right){\mathbf{E}}} \right]} 
\end{array}} \\ 
  {{\mathbf{X}} \in {{\left\{ {0,1} \right\}}^{nk}}\begin{array}{*{20}{c}}
  ,&{{{\mathbf{Q}}_{\mathbf{I}}} = {{\mathbf{A}}^T}{\mathbf{A}}} 
\end{array}} 
\end{array}
\end{equation}

\subsection{Parallel Augmentations with Limited Elements of the Graver Bases}
From a feasible solution, we augment using the Graver basis. Depending on where the initial feasible solution is and where the unknown optimal solution falls, the path between these two points is a vector that falls into a specific orthant. The sooner the algorithm derives the Graver basis inside this unknown orthant, the quicker it reaches the optimal solution. Thus, it is better if one starts with several feasible solutions and perform augmentation on several of them, in parallel. Thus, during the iterative course of building Graver basis, there would be a higher likelihood of reaching the optimum value faster. The integer solution can be recovered from the binary solution after each run using the integer to binary transformation given in equation (\ref{eq:encoding}).
         
\subsection{Numerical Results}\label{subsec:optexperiments}
We explore two different nonlinear integer programs. In the first case (motivated by a problem in Finance), the integer variables are binary. Therefore, $l = {\mathbf{0}}$ and $u = {\mathbf{1}}$,  so the required Graver basis is a truncated (boxed) Graver basis $\mathcal{G}(A), g_i \in {\left\{ { - 1,0, 1} \right\}^n}$. The second example
is an integer optimization problem with a low integer span $x \in {\left\{ { - 2,..., 2} \right\}^n}$.

~~\\
In both examples, we have used Algorithm \ref{alg:GE} to generate a truncated Graver basis and Algorithm \ref{alg:GOA} to optimize the given cost function.

\subsubsection{Case 1: Nonlinear 0-1 program with 50 binary variables}
The \textit{risk-averse capital budgeting problem} with integer (or binary) variables is of the form~\cite{atamturk_polymatroids_2008}:
\begin{equation}\label{eq:finance}
\left\{ {\begin{array}{*{20}{c}}
  {\begin{array}{*{20}{c}}
  {\min }&{ - \sum\nolimits_{i = 1}^n {{\mu _i}{x_i} + \sqrt {\frac{{1 - \varepsilon }}{\varepsilon }\sum\nolimits_{i = 1}^n {\sigma _i^2{x_i}^2} } } } 
\end{array}} \\ 
  {\begin{array}{*{20}{c}}
  {Ax = b}&,&{x \in {{\{ 0,1\} }^n}} 
\end{array}} 
\end{array}} \right.
\end{equation}
where ${\mu _i}'s$ are expected return values, ${\sigma _i}'s$ are variances, and $\varepsilon > 0$ represents the level of risk an investor is willing to take. The linear constraints are the risk-averse limits on the capital budget. The solution(s) to this problem are efficient portfolios, which matches the budget constraints. The objective function is a sum of a linear and a convex nonlinear term.

~~\\ 
We randomly generated many $(5 \times 50)$ binary matrices $A$.
Similar to \cite{atamturk_polymatroids_2008}, we set $b$ as close as half of row-sum of $A$. Also, ${\mu _i}$ is chosen from the uniform distribution $[0,1]$ and ${\sigma _i}$ is chosen from the uniform distribution $[0,\mu_i]$. The risk factor is set at $\varepsilon = 0.01$ to make the problem realistic as well as among the hardest to solve for classical MIP solvers.
~~\\
% Our goal is to compare and contrast solutions, using our Graver based Algorithm \ref{alg:GOA}, versus MIP solver of \textit{Gurobi Optimizer 8.1 \textsuperscript{TM}} and report the results. 
~~\\
For  $n=50$, we needed at least $k_i=2$ bits to cover $\{-1,0,1\}$ span needed for  a truncated Graver basis. This makes the problem size equal to $100$, which is not embeddable on the current D-Wave chip. But if we choose $k_i=1$ bit, by benefiting from suboptimal post processing, the span of results doubles (as we   discussed   in \ref{subsec:suboptimal}). All embeddings in this problem are a complete graph embedding of size $50$, which is very convenient. For $k_i=1$, which is the combinatorial case, the $2^i$ exponential binary encoding terms do not exist (except $2^0=1)$; therefore, $q_{ij}$ or $J_{ij}$ coupler values stay more uniform. Also, having only $0$ and $1$ in integer matrix $A$ results in fewer number of (unique) coupler values. These factors cause fewer chain breaks;  therefore, chainbreak strategy \textit{majority vote} works as well as \textit{minimize energy}, in this specific configuration. Both yield many unique kernels, which help to extract  Graver basis elements efficiently.

~~\\
%\textbf{Example 1:}
Here is an example that typifies several of our findings.
The matrix is:
\begin{equation}
A =
\left(\begin{array}{cccccccccccccccccccccccccccccccccccccccccccccccccc} 0 & 0 & 0 & 0 & 0 & 1 & 0 & 0 & 1 & 0 & 1 & 0 & 1 & 0 & 1 & 1 & 1 & 0 & 0 & 1 & 0 & 0 & 0 & 0 & 0 & 1 & 1 & 0 & 0 & 1 & 1 & 1 & 1 & 0 & 1 & 0 & 0 & 0 & 1 & 0 & 0 & 1 & 0 & 0 & 0 & 1 & 1 & 1 & 0 & 0\\ 0 & 0 & 0 & 0 & 1 & 0 & 0 & 0 & 1 & 0 & 0 & 1 & 0 & 1 & 1 & 0 & 1 & 0 & 0 & 1 & 0 & 0 & 0 & 1 & 1 & 1 & 0 & 1 & 0 & 1 & 1 & 1 & 1 & 1 & 1 & 0 & 0 & 0 & 0 & 1 & 1 & 0 & 0 & 0 & 1 & 0 & 0 & 1 & 0 & 0\\ 0 & 0 & 0 & 0 & 0 & 1 & 0 & 1 & 0 & 1 & 1 & 0 & 0 & 1 & 1 & 1 & 0 & 1 & 1 & 1 & 0 & 0 & 0 & 0 & 1 & 0 & 1 & 1 & 0 & 0 & 0 & 0 & 1 & 0 & 0 & 1 & 0 & 1 & 0 & 0 & 0 & 0 & 1 & 1 & 0 & 1 & 0 & 0 & 1 & 0\\ 1 & 1 & 1 & 1 & 1 & 0 & 0 & 1 & 0 & 0 & 1 & 1 & 1 & 0 & 0 & 1 & 0 & 1 & 1 & 1 & 0 & 1 & 0 & 1 & 0 & 0 & 1 & 1 & 0 & 1 & 1 & 0 & 1 & 0 & 0 & 0 & 1 & 0 & 0 & 1 & 1 & 0 & 0 & 1 & 0 & 1 & 0 & 0 & 1 & 0\\ 0 & 1 & 0 & 0 & 0 & 1 & 0 & 1 & 1 & 0 & 1 & 0 & 0 & 1 & 0 & 1 & 0 & 1 & 1 & 0 & 0 & 0 & 0 & 0 & 1 & 0 & 0 & 0 & 0 & 0 & 1 & 0 & 0 & 1 & 0 & 1 & 1 & 1 & 1 & 0 & 1 & 1 & 1 & 0 & 0 & 0 & 1 & 0 & 0 & 1 \end{array}\right)\nonumber
\end{equation}
The expected return vector is also created as a uniform real random variable in $[0,1]$ range:
\begin{equation}
\mu = 
\left(\begin{array}{ccccccccccc} 0.4718 & 0.7759 & 0.8211 & 0.2811 & 0.6575 & ... & 0.5669 & 0.1867 & 0.406 & 0.3712 & 0.09876 \end{array}\right)^T
\nonumber
\end{equation}
The variance for each term $\sigma_i$ is a real random variable between 0 and $\mu_i$:
\begin{equation}
\sigma = 
\left(\begin{array}{ccccccccccc} 0.4383 & 0.5841 & 0.07405 & 0.09704 & 0.1738 & ... & 0.337 & 0.1412 & 0.05004 & 0.3668 & 0.00677 \end{array}\right)^T
\nonumber
\end{equation}
We chose the budget values equal to half of the sum of each row of $A$:
$ b =
\left(\begin{array}{ccccc} 10 & 11 & 10 & 13 & 11 \end{array}\right)^T $ based on \cite{atamturk_polymatroids_2008}.
Readers can use this code\footnote{~~\\rng = 1; m = 5; n = 50;
A = randi([0 1],m,n);
mu = rand(n,1);
sigma = rand(n,1); sigma = sigma .* mu;
b = round(sum(A')/2)';}
in Matlab to regenerate the exact problem values $(A, \mu, \sigma, b)$. Also, $\varepsilon = 0.01$ to make the problem harder to solve, according to \cite{atamturk_polymatroids_2008}.

~~\\
The Graver basis of matrix $A$ inside the box $\left\{ { - {\mathbf{1}},..,{\mathbf{1}}} \right\}$
has $304$ ($608$ including symmetric elements) basis elements. The first and last $10$ vectors of the Graver basis are shown in Figure \ref{fig:graversize50}. This Graver basis is calculated using one call of D-Wave ($10000$ reads, annealing time = $1\mu sec.$) using \textit{minimize energy} chainbreak strategy (or three calls using \textit{majority vote} chainbreak strategy).

~~\\
We solved the $Ax = b$ using one call to the D-Wave quantum annealer, with $10000$ reads, annealing time = $1\mu sec.$. Over $6000$ unique solutions were obtained using \textit{minimize energy} chainbreak (and over $3000$  unique solutions using \textit{majority vote} chainbreak strategy).

~~\\
Evaluating the optimization cost function over the set of all obtained feasible solutions gives us a range of cost values and is shown in Figure \ref{fig:costvalues-1}. What is interesting -- as we see after augmentation -- is that the optimal solution is quite far way from these 6000+ solutions.
\begin{figure}[ht]
\centering
\includegraphics[width=10cm]{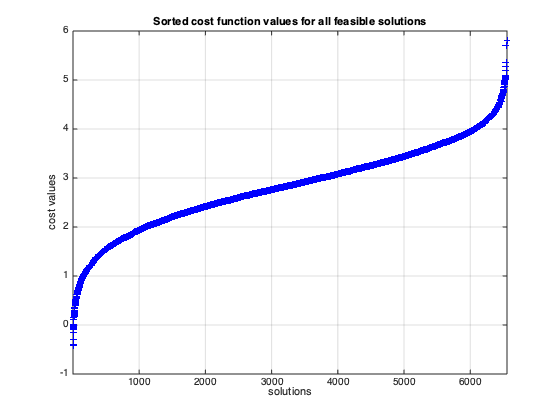}
\caption{The sorted cost values of all feasible solutions}
\label{fig:costvalues-1}
\end{figure}
~~\\
% \newpage
\begin{figure}
\begin{equation}
\mathcal G (A) =
\left(\begin{array}{ccccccccccccccccccccc} 0 & 0 & 0 & 0 & 0 & 0 & 0 & 0 & 0 & 0 & ... & 1 & 1 & 1 & 1 & 1 & 1 & 1 & 1 & 1 & 1\\ 0 & 0 & 0 & 0 & 0 & 0 & 0 & 0 & 0 & 0 & ... & 0 & 0 & 0 & 0 & 0 & 0 & 0 & 0 & 0 & 0\\ 0 & 0 & 0 & 0 & 0 & 0 & 0 & 0 & 0 & 0 & ... & 0 & 0 & 0 & 0 & 0 & 0 & 0 & 0 & 0 & 0\\ 0 & 0 & 0 & 0 & 0 & 0 & 0 & 0 & 0 & 0 & ... & 0 & 0 & 0 & 0 & 0 & 0 & 0 & 0 & 0 & 0\\ 0 & 0 & 0 & 0 & 0 & 0 & 0 & 0 & 0 & 0 & ... & 0 & 0 & 0 & 0 & 0 & 0 & 0 & 0 & 0 & 0\\ 0 & 0 & 0 & 0 & 0 & 0 & 0 & 0 & 0 & 0 & ... & 0 & 0 & 0 & 0 & 0 & 0 & 0 & 0 & 0 & 0\\ 0 & 0 & 0 & 0 & 0 & 0 & 0 & 0 & 0 & 0 & ... & 0 & 0 & 0 & 0 & 0 & 0 & 0 & 0 & 0 & 0\\ 0 & 0 & 0 & 0 & 0 & 0 & 0 & 0 & 0 & 0 & ... & 0 & 0 & 0 & 0 & 0 & 0 & 0 & 0 & 0 & 0\\ 0 & 0 & 0 & 0 & 0 & 0 & 0 & 0 & 0 & 0 & ... & 0 & 0 & 0 & 0 & 0 & 0 & 0 & 0 & 0 & 0\\ 0 & 0 & 0 & 0 & 0 & 0 & 0 & 0 & 0 & 0 & ... & 0 & 0 & 0 & 0 & 0 & 1 & 1 & 1 & 1 & 1\\ 0 & 0 & 0 & 0 & 0 & 0 & 0 & 0 & 0 & 0 & ... & 0 & 0 & 0 & 0 & 0 & 0 & 0 & 0 & 0 & 0\\ 0 & 0 & 0 & 0 & 0 & 0 & 0 & 0 & 0 & 0 & ... & 0 & 0 & 0 & 0 & 0 & 0 & 0 & 0 & 0 & 0\\ 0 & 0 & 0 & 0 & 0 & 0 & 0 & 0 & 0 & 0 & ... & 0 & 0 & 0 & 0 & 0 & 0 & 0 & 0 & 0 & 0\\ 0 & 0 & 0 & 0 & 0 & 0 & 0 & 0 & 0 & 0 & ... & 0 & 0 & 0 & 0 & 0 & 0 & 0 & 0 & 0 & 0\\ 0 & 0 & 0 & 0 & 0 & 0 & 0 & 0 & 0 & 0 & ... & 0 & 0 & 0 & 0 & 0 & 0 & 0 & 0 & 0 & 0\\ 0 & 0 & 0 & 0 & 0 & 0 & 0 & 0 & 0 & 0 & ... & 0 & 0 & 0 & 0 & 0 & 0 & 0 & 0 & 0 & 0\\ 0 & 0 & 0 & 0 & 0 & 0 & 0 & 0 & 0 & 0 & ... & 0 & 0 & 0 & 0 & 0 & 0 & 0 & 0 & 0 & 0\\ 0 & 0 & 0 & 0 & 0 & 0 & 0 & 0 & 0 & 0 & ... & 0 & 0 & 0 & 0 & 0 & -1 & 0 & 0 & 0 & 0\\ 0 & 0 & 0 & 0 & 0 & 0 & 0 & 0 & 0 & 0 & ... & 0 & 0 & 0 & 0 & 0 & 0 & -1 & 0 & 0 & 0\\ 0 & 0 & 0 & 0 & 0 & 0 & 0 & 0 & 0 & 0 & ... & 0 & 0 & 0 & 0 & 0 & 0 & 0 & 0 & 0 & 0\\ 0 & 0 & 0 & 0 & 0 & 0 & 0 & 0 & 0 & 0 & ... & 0 & 0 & 0 & 0 & 0 & 0 & 0 & 0 & 0 & 0\\ 0 & 0 & 0 & 0 & 0 & 0 & 0 & 0 & 0 & 0 & ... & 0 & 0 & 0 & 0 & 0 & 0 & 0 & 0 & 0 & 0\\ 0 & 0 & 0 & 0 & 0 & 0 & 0 & 0 & 0 & 0 & ... & 0 & 0 & 0 & 0 & 0 & 0 & 0 & 0 & 0 & 0\\ 0 & 0 & 0 & 0 & 0 & 0 & 0 & 0 & 0 & 0 & ... & 0 & 0 & 0 & 0 & 0 & 0 & 0 & 0 & 0 & 0\\ 0 & 0 & 0 & 0 & 0 & 0 & 0 & 0 & 0 & 0 & ... & 0 & 0 & 0 & 0 & 0 & 0 & 0 & 0 & 0 & 0\\ 0 & 0 & 0 & 0 & 0 & 0 & 0 & 0 & 0 & 0 & ... & 0 & 0 & 0 & 0 & 1 & 0 & 0 & 0 & 0 & 0\\ 0 & 0 & 0 & 0 & 0 & 0 & 0 & 0 & 0 & 0 & ... & 0 & 0 & 0 & 0 & 0 & 0 & 0 & 0 & 0 & 0\\ 0 & 0 & 0 & 0 & 0 & 0 & 0 & 0 & 0 & 0 & ... & 0 & 0 & 0 & 0 & 0 & 0 & 0 & -1 & 0 & 0\\ 0 & 0 & 0 & 0 & 0 & 0 & 0 & 0 & 0 & 0 & ... & 0 & 0 & 0 & 0 & 0 & 0 & 0 & 0 & 0 & 0\\ 0 & 0 & 0 & 0 & 0 & 0 & 0 & 0 & 0 & 0 & ... & 0 & 0 & 0 & 0 & -1 & 0 & 0 & 0 & 0 & 0\\ 0 & 0 & 0 & 0 & 0 & 0 & 0 & 0 & 0 & 0 & ... & 0 & 0 & 0 & 0 & 0 & 0 & 0 & 0 & 0 & 0\\ 0 & 0 & 0 & 0 & 0 & 0 & 0 & 0 & 0 & 0 & ... & 0 & 0 & 0 & 0 & 0 & 0 & 0 & 0 & 0 & 0\\ 0 & 0 & 0 & 0 & 0 & 0 & 0 & 0 & 0 & 0 & ... & 0 & 0 & 0 & 0 & 0 & 0 & 0 & 0 & 0 & 0\\ 0 & 0 & 0 & 0 & 0 & 0 & 0 & 0 & 0 & 0 & ... & 0 & 0 & 1 & 1 & 0 & 0 & 0 & 0 & 0 & 0\\ 0 & 0 & 0 & 0 & 0 & 0 & 0 & 0 & 0 & 0 & ... & 0 & 0 & 0 & 0 & 0 & 0 & 0 & 0 & 0 & 0\\ 0 & 0 & 0 & 0 & 0 & 0 & 0 & 0 & 1 & 1 & ... & 1 & 1 & 0 & 0 & 0 & 0 & 0 & 0 & 0 & 0\\ 0 & 0 & 0 & 0 & 0 & 0 & 0 & 0 & 0 & 0 & ... & 0 & 0 & 0 & 0 & 0 & 0 & 0 & 0 & 0 & 0\\ 0 & 0 & 0 & 0 & 0 & 0 & 0 & 1 & -1 & 0 & ... & 0 & 0 & 0 & 0 & 0 & 0 & 0 & 0 & 0 & 0\\ 0 & 0 & 0 & 0 & 1 & 1 & 1 & 0 & 0 & 0 & ... & 0 & 0 & 0 & 0 & 0 & 0 & 0 & 0 & 0 & 0\\ 0 & 0 & 0 & 1 & 0 & 0 & 0 & 0 & 0 & 0 & ... & 0 & 0 & -1 & 0 & 0 & 0 & 0 & 0 & 0 & 0\\ 0 & 0 & 0 & -1 & 0 & 0 & 0 & 0 & 0 & 0 & ... & 0 & 0 & 0 & -1 & 0 & 0 & 0 & 0 & 0 & 0\\ 0 & 1 & 1 & 0 & -1 & 0 & 0 & 0 & 0 & 0 & ... & 0 & 0 & 0 & 0 & 0 & 0 & 0 & 0 & 0 & 0\\ 0 & 0 & 0 & 0 & 0 & 0 & 0 & -1 & 0 & -1 & ... & 0 & 0 & 0 & 0 & 0 & 0 & 0 & 0 & 0 & 0\\ 1 & 0 & 0 & 0 & 0 & 0 & 0 & 0 & 0 & 0 & ... & -1 & 0 & 0 & 0 & 0 & 0 & 0 & 0 & -1 & 0\\ 0 & 0 & 1 & 0 & 0 & 0 & 1 & 0 & 0 & 0 & ... & 0 & 0 & 0 & 0 & 0 & 0 & 0 & 1 & 0 & 0\\ 0 & 0 & 0 & 0 & 0 & 0 & 0 & 0 & 0 & 0 & ... & 0 & 0 & 0 & 0 & 0 & 0 & 0 & 0 & 0 & 0\\ 0 & -1 & 0 & 0 & 0 & -1 & 0 & 0 & 0 & 0 & ... & 0 & 0 & 0 & 0 & 0 & 0 & 0 & 0 & 0 & 0\\ 0 & 0 & -1 & 0 & 0 & 0 & -1 & 0 & 0 & 0 & ... & 0 & 0 & 0 & 0 & 0 & 0 & 0 & 0 & 0 & 0\\ -1 & 0 & 0 & 0 & 0 & 0 & 0 & 0 & 0 & 0 & ... & 0 & -1 & 0 & 0 & 0 & 0 & 0 & 0 & 0 & -1\\ 0 & 0 & -1 & 1 & 0 & 0 & -1 & 0 & 0 & 0 & ... & -1 & -1 & -1 & 0 & 0 & 1 & 1 & 0 & 0 & 0 \end{array}\right)
\nonumber
\end{equation}
\caption{The first and last 10 elements of the Graver basis of the matrix $A$ of Example 1}
\label{fig:graversize50}
\end{figure}
 
 ~~\\
 As it can be seen, for various solutions the cost values changes from a maximum of $5.7999$ to a minimum of $-0.4154$. Clearly, a wise choice of initial feasible solution is to choose the minimum solution corresponding to the minimum cost value, which is,
 \begin{equation}
 x_0 = 
 \left(\begin{array}{cccccccccccccccccccccccccccccccccccccccccccccccccc} 0 & 0 & 1 & 1 & 1 & 1 & 1 & 0 & 0 & 0 & 0 & 0 & 1 & 0 & 0 & 1 & 0 & 1 & 1 & 1 & 1 & 0 & 1 & 0 & 1 & 0 & 0 & 1 & 1 & 0 & 0 & 1 & 1 & 1 & 1 & 0 & 0 & 1 & 0 & 1 & 1 & 1 & 0 & 0 & 0 & 0 & 1 & 1 & 1 & 1 \end{array}\right)^T
 \nonumber
\end{equation}
If we start the augmentation procedure from the minimum $x_0$, using $24$ augmenting (improving) steps, we reach the global solution of the problem, which has a value of $-3.6947$. A  global solution for this problem is,
\begin{equation}
x^* =
\left(\begin{array}{cccccccccccccccccccccccccccccccccccccccccccccccccc} 0 & 0 & 1 & 1 & 1 & 1 & 1 & 0 & 1 & 0 & 0 & 0 & 1 & 0 & 0 & 1 & 0 & 1 & 1 & 1 & 1 & 0 & 1 & 1 & 0 & 0 & 1 & 1 & 1 & 0 & 0 & 0 & 1 & 1 & 1 & 0 & 0 & 1 & 1 & 0 & 1 & 0 & 1 & 0 & 1 & 0 & 0 & 1 & 0 & 1 \end{array}\right)^T
\nonumber
\end{equation}
~~\\
If we instead start the augmentation from the farthest feasible point among all feasible points, which is the solution corresponding to the cost value $5.7999$:
\begin{equation}
x_0 =
\left(\begin{array}{cccccccccccccccccccccccccccccccccccccccccccccccccc} 1 & 1 & 0 & 0 & 1 & 0 & 1 & 1 & 1 & 1 & 1 & 1 & 0 & 1 & 1 & 1 & 1 & 0 & 0 & 0 & 1 & 0 & 1 & 1 & 1 & 1 & 0 & 0 & 1 & 1 & 1 & 0 & 0 & 0 & 0 & 1 & 1 & 0 & 0 & 0 & 0 & 1 & 0 & 0 & 0 & 1 & 0 & 0 & 1 & 0 \end{array}\right)^T
\nonumber
\end{equation}
using $30$ augmenting (improving) steps we reach the same global solution.

~~\\
As  we can observe, although we initially had over $6000$ feasible points, the minimum cost value among these was $-0.4154$ which is not optimal. From the cost value of $-0.4154$ to the optimum cost ($-3.6947$), we needed $24$ augmentations. This does not mean that we only did $24$ comparisons. In fact, we tested all of the $ 2 \times 304$ Graver basis elements, three times. Many of the choices of the form $x+g$ resulted in values outside $\{0,1\}$ and could be immediately ignored before calculating their cost.
\subsubsection{Comparison with a best-in-class classical method}
 For a quick comparison with a leading edge classical solver, the same problem was passed to the Gurobi Optimizer (latest version, 8.0)\footnote{Installed on MacBookPro15,1: 6 Core 2.6 GHz Intel Core i7 processor, 32 GB 2400 MHz DDR4 RAM}.
Gurobi solved the above problem in less than $0.2 sec$, which is much faster than our algorithm on D-Wave Processor. If we increase the span of random integer terms ($a_{ij}$'s) in matrix $A$, from $\{0,1\}$ to $\{0, ...,t\}$, by increasing $t$, the problem becomes harder and harder for Gurobi. At $t=10$, Gurobi needs $16sec$, at $t=20$ over $3min$, at $t=30$ over $5min$, at $t=40$ over $21min$, at $t=50$ over $75min$, ... , and at $t=100,$ even after 8 hours Gurobi could not solve the problem. Note that for our quantum annealing algorithm, the size and number of calls to D-Wave and the embedding for all of those $t$  stays exactly the same. The only difference is that with increasing $t$, the number of unique terms in the QUBO $q_{ij}$'s that maps to the Ising model $J_{ij}$'s (i.e. the cardinality of set of $J_{ij}$ values) increases.

~~\\
For $t=1$, there are $6$ unique terms, for $t=10$: $\simeq 230$, for $t=20$: $\simeq 620$, for $t=30$: $\simeq 860$, for $t=40$: $\simeq 1030$, for $t=50$: $\simeq 1070$, and for $t=100$: $\simeq 1190$ unique terms required to be mapped into a coupler.
The current hardware has a limited precision ($4bits = 16$ steps) to address coupler instances. This  matches the case of $t=1$. Therefore, if the next generation of quantum annealers invests in improving \textit{only} the coupler precision (without increasing the number of qubits or improving connectivity between them), we can match, or even surpass,  the current classical solvers, using our quantum algorithm, in a wide range of nonlinear integer programming problems.

~~\\
Now if we fix the matrix parameters (e.g. $t=1$) and increase problem size ($n$), solving the problem still becomes harder for Gurobi solver, whilst the number of unique coupler terms roughly stays the same. 
For random problems with $A^{20 \times 80}$ (size $80$, $12$ unique terms), Gurobi needs $\sim 135sec$, for $A^{25 \times 100}$ (size $100$, $16$ terms), about $\sim 2hours$ and for $A^{30 \times 120}$ (size $120$, $16$ terms), even after $3hours$ Gurobi did not converge to an optimal solution.

~~\\
D-Wave's technical report \cite{boothby_next-generation_2018} about their next generation chip - due to newly designed denser connectivity structure called \textit{Pegasus}- indicates it can embed complete graph problems of size up to 180, and do so with lower chain lengths than the current D-Wave 2000Q processor. We estimate, even if the new processor's coupler precision is increased by only one bit, due to embedding size expansion, our quantum classical algorithm can surpass the speed of best-in-class classical solvers (e.g. Gurobi), in similar nonlinear combinatorial programming problems with sizes greater that $100$, up until $180$.

~~\\
Our approach is a first that benefits from many anneals (thus reads) of quantum annealer in each call, not to increase the confidence of \textit{an optimal solution}, but instead to sample \textit{many optimal and near-optimal solutions} of highly degenerate problems. Many unique feasible solutions can be found very quickly. If
the next generation of quantum annealing processors can increase the maximum number of anneals per each call (from $10000$ to say $100000$), it will significantly enhance the performance of our approach as the number of quantum calls needed can be just one for Graver, and another for obtaining many good starting points for optimization.  

\subsubsection{Case 2: Nonlinear low span integer program with 25 integer variables}
Consider again the nonlinear programming problem shown in Equation (\ref{eq:finance}), but now in bounded integer form ${x \in {{\{ -2, -1, 0, 1, 2\} }^n}}$.
% Integer variables span in a limited range, here $\{-2,-1,0, 1, 2\}$. 
To be able to embed the problem into the current processor, we generate a $5 \times 25$ problem and use $k=2$ bits for each integer variable.

~~\\
Consider a typical example, where the matrix is:

$$ A = 
\left(\begin{array}{ccccccccccccccccccccccccc} 1 & 1 & 1 & 1 & 1 & 1 & 1 & 1 & 1 & 1 & 0 & 1 & 0 & 1 & 0 & 1 & 0 & 1 & 1 & 1 & 0 & 1 & 0 & 1 & 0\\ 1 & 1 & 1 & 1 & 0 & 1 & 0 & 1 & 0 & 0 & 1 & 0 & 0 & 0 & 1 & 0 & 0 & 1 & 0 & 1 & 1 & 1 & 1 & 1 & 1\\ 0 & 1 & 0 & 0 & 0 & 1 & 0 & 1 & 0 & 1 & 1 & 0 & 1 & 1 & 0 & 1 & 1 & 0 & 0 & 1 & 0 & 0 & 1 & 1 & 1\\ 0 & 0 & 0 & 0 & 0 & 0 & 0 & 1 & 0 & 1 & 1 & 1 & 0 & 1 & 1 & 1 & 1 & 0 & 0 & 1 & 0 & 0 & 0 & 0 & 0\\ 0 & 1 & 1 & 1 & 1 & 1 & 0 & 0 & 0 & 1 & 0 & 0 & 0 & 1 & 0 & 0 & 0 & 0 & 0 & 1 & 0 & 1 & 0 & 1 & 0 \end{array}\right)
$$
~~\\
The expected return vector is a uniform real random variable of size $25$ in $[0,1]$ range:
\begin{equation}
\mu = 
\left(\begin{array}{ccccccccccc} 0.9277 & 0.06858 & 0.2994 & 0.5916 & 0.2033 & ... & 0.5098 & 0.9742 & 0.1973 & 0.1112 & 0.2974 \end{array}\right)^T
\nonumber
\end{equation}
The variance for each term $\sigma_i$ is a real random variable between 0 and $\mu_i$:
\begin{equation}
\sigma = 
\left(\begin{array}{ccccccccccc} 0.3677 & 0.02886 & 0.09326 & 0.4105 & 0.01868 & ... & 0.4081 & 0.8942 & 0.02709 & 0.05612 & 0.1204 \end{array}\right)^T
\nonumber
\end{equation}
We chose the budget values equal to half of sum of each row of $A$:
$ b =
\left(\begin{array}{ccccc} 9 & 8 & 7 & 5 & 5 \end{array}\right)^T $ based on \cite{atamturk_polymatroids_2008}.

~~\\
The lower and upper bounds for the problem variables are $l=-2\textbf{1}$ and $u=+2\textbf{1}$. Therefore, the truncated Graver basis bound is $[L_g,U_g] = [-4 \textbf{1}, +4 \textbf{1}]$.
Spanning the variables in this range needs a little more than the average $3$ bits encoding. We calculated all of the Graver basis elements in this range using only $2$ bits encoding length, in two D-Wave calls with a center adjustment and benefiting from the range doubling obtained from suboptimal post-processings (see \ref{subsec:suboptimal}).
Matrix $A$ in this band contains $2 \times 616$ Graver elements.

~~\\
Separately, we solved   $Ax = b$ using one call to the D-Wave quantum annealer, having only two bits of encoding ($10000$ reads, annealing time = $1\mu sec.$). We chose the center of the two bit encoding at $0$. Therefore, it covers $\{-2, -1, 0, 1\}$, which is enough for finding an initial point. The number of unique optimal solutions was $773$  using \textit{minimize energy} chainbreak and $238$   using \textit{majority vote} chainbreak.

~~\\
Evaluating the optimization cost function over the set of all solutions gives us a range of cost values, as shown below.

\begin{figure}[ht]
\centering
\includegraphics[width=10cm]{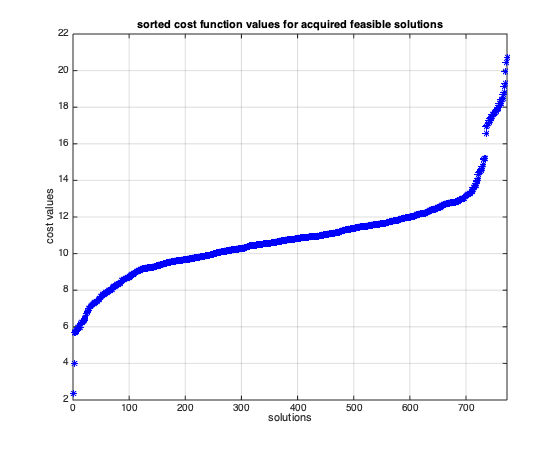}
\caption{The sorted cost values of acquired feasible solutions for low span example}
\label{fig:costvalues-2}
\end{figure}
~~\\
The cost values range from maximum of $20.7211$ for the initial feasible solution:
$$
x_0^{max} = 
\left(\begin{array}{ccccccccccccccccccccccccc} 1 & 1 & 1 & 1 & -1 & 1 & 1 & 1 & 1 & 1 & 1 & 1 & -2 & 1 & 0 & -1 & 0 & 1 & -1 & 1 & -2 & -2 & 1 & 1 & 1 \end{array}\right)^T
$$
to minimum of $2.3622$ for the initial feasible solution:
$$
x_0^{min} = 
\left(\begin{array}{ccccccccccccccccccccccccc} 0 & 1 & 0 & 1 & 1 & 1 & 1 & 0 & 1 & 1 & 1 & 0 & 0 & 1 & 1 & 0 & 1 & 1 & 1 & 0 & 1 & 0 & 1 & -1 & 1 \end{array}\right)^T
$$
~~\\
Starting from $x_0^{max}$ using $22$ augmenting (improving) steps, the global solution is achieved, while starting from $x_0^{min}$ using only $8$ augmenting steps, the same global solution is achieved:
$$
x^{*} = 
\left(\begin{array}{ccccccccccccccccccccccccc} 0 & 1 & 1 & 0 & 2 & 0 & 0 & 0 & 2 & 0 & 2 & 1 & -2 & 0 & 0 & 0 & 2 & 0 & 1 & 0 & 0 & 0 & 2 & 1 & 1 \end{array}\right)^T
$$
which has a minimum cost of $-2.4603$.

\subsubsection{Augmenting with partial Graver basis}
In the above, we assumed that all of the truncated Graver basis is acquired, thus starting from any of the many available initial solutions, we reach the global solution. In fact, the starting point affects only the number of augmentations required to reach the final solution. Theoretically, if we have only one feasible solution and all of the Graver basis elements in the required region, we certainly reach the global solution in polynomial number of augmentations, for the proved class of nonlinear costs (see \ref{subsec:costcategories}).
Looking from the other side of spectrum, if we have all feasible solutions and none of the Graver basis elements, just by checking each feasible solution's cost and finding the minimum among them, we reach the global solution.

~~\\
Our quantum annealing-based algorithm provides a third opportunity. If we have many feasible solutions distributed in the solution space (let us say, uniformly), then we \textit{most likely} may reach the global solution even with a subset of Graver basis elements (which we refer to as partial Graver basis). The point is, if we can not reach from a particular feasible solution to a global one, due to lack of   \textit{specific} Graver basis elements in the available set, other feasible solutions provide many other paths (to the global solution) that may not need these \textit{specific} Graver basis elements that are missing.

~~\\
This interesting opportunity has borne out in our experiments. 
In the previous example, if we  call D-Wave only once for the Graver basis, we acquire almost $2 \times 418$ Graver basis elements, which is $68\%$ of the Graver basis elements previously found. Now, we start augmenting the problem from all of $773$ initial feasible points. Interestingly, using this partial Graver basis, we could reach the global solution from $64$ (out of $773$) of the initial solutions; see Figure (\ref{fig:costvalues-augp}) The other paths also terminate to very good near-optimal solutions (compare average values in Figure \ref{fig:costvalues-2}, $10.9013$, versus average values in Figure \ref{fig:costvalues-augp}, $-1.6546$).

~~\\
Furthermore, in case there is degeneracy in the global solution (that is, there are alternate optimal solutions), this multi augmentation strategy can also result in several unique degenerate solutions. In the above example, the problem does not have a degenerate solution, and so all $64$ paths resulted in the same solution $x^*$, the only one.

~~\\
Using many or all of the feasible solutions clearly adds linear computational cost to augmentation, but the augmentation cost is low (polynomial) compared with the Graver cost.
Finding a good balance between the number of feasible solutions and the amount and variability of Graver basis elements would be an interesting experimental study.
~~\\

\begin{figure}[H]
\centering
\includegraphics[width=10cm]{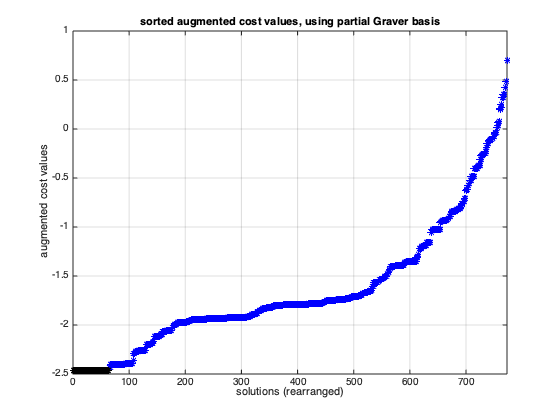}
\caption{The sorted augmented cost values for  all feasible solutions, using partially acquired Graver basis}
\label{fig:costvalues-augp}
\end{figure}

\section{Conclusions}\label{sec:conclusions}
In this paper, we demonstrate that calculation of Graver bases using quantum annealing, and its subsequent use for nonlinear integer optimization, is a viable approach in available hardware. To accomplish this, we developed several post-processing procedures, and an adaptive approach that adjusts the encoding length as well as the center of the variables to be efficient in the use of the limited number of qubits.

~~\\
We did not expect to outperform best-in-class classical solvers, and indeed that was not our motivation. If anything, we began as skeptics about the practical capabilities of the available hardware. We are pleasantly surprised that we are able to solve many problems of medium size in decent time, not too far behind the classical best, if we select problem instances that are hard for classical approaches. Our tests suggest that with improvements in the number of qubits and their connectivity, which have already been  announced and are expected in the near term, an improvement in the coupler precision will bring our hybrid quantum-classical approach within reach of, or even surpass, the best-in-class classical algorithms for an important class of non-linear integer optimization problems. 

\section*{Acknowledgements}
\addcontentsline{toc}{section}{Acknowledgements}
The authors thank Davide Venturelli, Eleanor G. Rieffel, Bryan O’Gorman, and other NASA QuAIL team members for their comments and feedback.

\section*{Bibliography}
\addcontentsline{toc}{section}{Bibliography}\label{sec:bibliography}
\bibliographystyle{amsalpha}
\bibliography{main}

\newcommand{\etalchar}[1]{$^{#1}$}
\providecommand{\bysame}{\leavevmode\hbox to3em{\hrulefill}\thinspace}
\providecommand{\MR}{\relax\ifhmode\unskip\space\fi MR }
\providecommand{\MRhref}[2]{%
  \href{http://www.ams.org/mathscinet-getitem?mr=#1}{#2}
}
\providecommand{\href}[2]{#2}
\begin{thebibliography}{DLHO{\etalchar{+}}09}

\bibitem[AN08]{atamturk_polymatroids_2008}
Alper Atamtürk and Vishnu Narayanan, \emph{Polymatroids and mean-risk
  minimization in discrete optimization}, Operations Research Letters
  \textbf{36} (2008), no.~5, 618--622.

\bibitem[Bar82]{barahona_computational_1982}
F.~Barahona, \emph{On the computational complexity of {Ising} spin glass
  models}, Journal of Physics A: Mathematical and General \textbf{15} (1982),
  no.~10, 3241 (en).

\bibitem[Ber18]{berwald_mathematics_2018}
Jesse~J. Berwald, \emph{The {Mathematics} of {Quantum}-{Enabled} {Applications}
  on the {D}-{Wave} {Quantum} {Computer}}, arXiv:1812.00062 [quant-ph] (2018),
  arXiv: 1812.00062.

\bibitem[BHJ{\etalchar{+}}14]{bunyk_architectural_2014}
P.~I. Bunyk, E.~M. Hoskinson, M.~W. Johnson, E.~Tolkacheva, F.~Altomare, A.~J.
  Berkley, R.~Harris, J.~P. Hilton, T.~Lanting, A.~J. Przybysz, and
  J.~Whittaker, \emph{Architectural {Considerations} in the {Design} of a
  {Superconducting} {Quantum} {Annealing} {Processor}}, IEEE Transactions on
  Applied Superconductivity \textbf{24} (2014), no.~4, 1--10.

\bibitem[BLSR99]{bigatti_computing_1999}
A.~M. Bigatti, R.~La~Scala, and L.~Robbiano, \emph{Computing {Toric} {Ideals}},
  Journal of Symbolic Computation \textbf{27} (1999), no.~4, 351--365.

\bibitem[Boo18]{boothby_next-generation_2018}
Kelly Boothby, \emph{Next-{Generation} {Topology} of {D}-{Wave} {Quantum}
  {Processors}}, D-Wave Technical Report (2018).

\bibitem[BPT00]{bertsimas_new_2000}
Dimitris Bertsimas, Georgia Perakis, and Sridhar Tayur, \emph{A {New}
  {Algebraic} {Geometry} {Algorithm} for {Integer} {Programming}}, Management
  Science \textbf{46} (2000), no.~7, 999--1008.

\bibitem[Cho08]{choi_minor-embedding_2008}
Vicky Choi, \emph{Minor-embedding in {Adiabatic} {Quantum} {Computation}: {I}.
  {The} {Parameter} {Setting} {Problem}}, Quantum Information Processing
  \textbf{7} (2008), no.~5, 193--209.

\bibitem[CT91]{conti_buchberger_1991}
Pasqualina Conti and Carlo Traverso, \emph{Buchberger {Algorithm} and {Integer}
  {Programming}}, Proceedings of the 9th {International} {Symposium}, on
  {Applied} {Algebra}, {Algebraic} {Algorithms} and {Error}-{Correcting}
  {Codes} (London, UK, UK), {AAECC}-9, Springer-Verlag, 1991, pp.~130--139.

\bibitem[CUWW97]{cornuejols_decomposition_1997}
G.~Cornuejols, R.~Urbaniak, R.~Weismantel, and L.~Wolsey, \emph{Decomposition
  of integer programs and of generating sets}, Algorithms — {ESA} '97 (Rainer
  Burkard and Gerhard Woeginger, eds.), Lecture {Notes} in {Computer}
  {Science}, Springer Berlin Heidelberg, 1997, pp.~92--103 (en).

\bibitem[DAT18a]{cmumt}
Raouf Dridi, Hedayat Alghassi, and Sridhar Tayur, \emph{Homological description
  of the quantum adiabatic evolution with a view toward quantum computations},
  arXiv:1811.00675 (2018).

\bibitem[DAT18b]{cmuag}
\bysame, \emph{A novel algebraic geometry compiling framework for adiabatic
  quantum computations}, arXiv:1810.01440 (2018).

\bibitem[DLHK12]{de_loera_algebraic_2012}
J.~De~Loera, R.~Hemmecke, and M.~Köppe, \emph{Algebraic and {Geometric}
  {Ideas} in the {Theory} of {Discrete} {Optimization}}, {MOS}-{SIAM} {Series}
  on {Optimization}, Society for Industrial and Applied Mathematics, December
  2012.

\bibitem[DLHO{\etalchar{+}}09]{de_loera_convex_2009}
J.~A. De~Loera, R.~Hemmecke, S.~Onn, U.~G. Rothblum, and R.~Weismantel,
  \emph{Convex integer maximization via {Graver} bases}, Journal of Pure and
  Applied Algebra \textbf{213} (2009), no.~8, 1569--1577.

\bibitem[FGGS00]{farhi_quantum_2000}
Edward Farhi, Jeffrey Goldstone, Sam Gutmann, and Michael Sipser, \emph{Quantum
  {Computation} by {Adiabatic} {Evolution}}, arXiv:quant-ph/0001106 (2000),
  arXiv: quant-ph/0001106.

\bibitem[Gra75]{graver_foundations_1975}
Jack~E. Graver, \emph{On the foundations of linear and integer linear
  programming {I}}, Mathematical Programming \textbf{9} (1975), no.~1, 207--226
  (en).

\bibitem[Hem02]{hemmecke_computation_2002}
Raymond Hemmecke, \emph{On the computation of hilbert bases of cones},
  Mathematical {Software}, World Scientific, July 2002, pp.~307--317.

\bibitem[HJL{\etalchar{+}}10]{harris_experimental_2010}
R.~Harris, M.~W. Johnson, T.~Lanting, A.~J. Berkley, J.~Johansson, P.~Bunyk,
  E.~Tolkacheva, E.~Ladizinsky, N.~Ladizinsky, T.~Oh, F.~Cioata, I.~Perminov,
  P.~Spear, C.~Enderud, C.~Rich, S.~Uchaikin, M.~C. Thom, E.~M. Chapple,
  J.~Wang, B.~Wilson, M.~H.~S. Amin, N.~Dickson, K.~Karimi, B.~Macready,
  C.~J.~S. Truncik, and G.~Rose, \emph{Experimental investigation of an
  eight-qubit unit cell in a superconducting optimization processor}, Physical
  Review B \textbf{82} (2010), no.~2, 024511.

\bibitem[HOW11]{hemmecke_polynomial_2011}
Raymond Hemmecke, Shmuel Onn, and Robert Weismantel, \emph{A polynomial
  oracle-time algorithm for convex integer minimization}, Mathematical
  Programming \textbf{126} (2011), no.~1, 97--117 (en).

\bibitem[HS95]{hosten_grin:_1995}
Serkan Hoşten and Bernd Sturmfels, \emph{{GRIN}: {An} implementation of
  {Gröbner} bases for integer programming}, Integer {Programming} and
  {Combinatorial} {Optimization}, Lecture {Notes} in {Computer} {Science},
  Springer, Berlin, Heidelberg, May 1995, pp.~267--276 (en).

\bibitem[JAG{\etalchar{+}}11]{johnson_quantum_2011}
M.~W. Johnson, M.~H.~S. Amin, S.~Gildert, T.~Lanting, F.~Hamze, N.~Dickson,
  R.~Harris, A.~J. Berkley, J.~Johansson, P.~Bunyk, E.~M. Chapple, C.~Enderud,
  J.~P. Hilton, K.~Karimi, E.~Ladizinsky, N.~Ladizinsky, T.~Oh, I.~Perminov,
  C.~Rich, M.~C. Thom, E.~Tolkacheva, C.~J.~S. Truncik, S.~Uchaikin, J.~Wang,
  B.~Wilson, and G.~Rose, \emph{Quantum annealing with manufactured spins},
  Nature \textbf{473} (2011), no.~7346, 194--198 (en).

\bibitem[Kat50]{kato_adiabatic_1950}
Tosio Kato, \emph{On the {Adiabatic} {Theorem} of {Quantum} {Mechanics}},
  Journal of the Physical Society of Japan \textbf{5} (1950), no.~6, 435--439.

\bibitem[LORW12]{lee_quadratic_2012}
Jon Lee, Shmuel Onn, Lyubov Romanchuk, and Robert Weismantel, \emph{The
  quadratic {Graver} cone, quadratic integer minimization, and extensions},
  Mathematical Programming \textbf{136} (2012), no.~2, 301--323 (en).

\bibitem[McG14]{mcgeoch_adiabatic_2014}
Catherine~C. McGeoch, \emph{Adiabatic {Quantum} {Computation} and {Quantum}
  {Annealing}: {Theory} and {Practice}}, Synthesis Lectures on Quantum
  Computing \textbf{5} (2014), no.~2, 1--93.

\bibitem[MSW04]{murota_optimality_2004}
Kazuo Murota, Hiroo Saito, and Robert Weismantel, \emph{Optimality criterion
  for a class of nonlinear integer programs}, Operations Research Letters
  \textbf{32} (2004), no.~5, 468--472.

\bibitem[MZK17]{mandra_exponentially_2017}
Salvatore Mandrà, Zheng Zhu, and Helmut~G. Katzgraber, \emph{Exponentially
  {Biased} {Ground}-{State} {Sampling} of {Quantum} {Annealing} {Machines} with
  {Transverse}-{Field} {Driving} {Hamiltonians}}, Physical Review Letters
  \textbf{118} (2017), no.~7, 070502.

\bibitem[noa]{noauthor_4ti2--software_nodate}
\emph{4ti2--{A} software package for algebraic, geometric and combinatorial
  problems on linear spaces}.

\bibitem[Onn10]{onn_nonlinear_2010}
Shmuel Onn, \emph{Nonlinear {Discrete} {Optimization}: {An} {Algorithmic}
  {Theory}}, European Mathematical Soc., 2010 (en), Google-Books-ID:
  kcD5DAEACAAJ.

\bibitem[Pap81]{papadimitriou_complexity_1981}
Christos~H. Papadimitriou, \emph{On the {Complexity} of {Integer}
  {Programming}}, J. ACM \textbf{28} (1981), no.~4, 765--768.

\bibitem[Pot96]{pottier_euclidean_1996}
Loïc Pottier, \emph{The {Euclidean} {Algorithm} in {Dimension} {N}},
  Proceedings of the 1996 {International} {Symposium} on {Symbolic} and
  {Algebraic} {Computation} (New York, NY, USA), {ISSAC} '96, ACM, 1996,
  pp.~40--42.

\bibitem[ST97]{sturmfels_variation_1997}
Bernd Sturmfels and Rekha~R. Thomas, \emph{Variation of cost functions in
  integer programming}, Mathematical Programming \textbf{77} (1997), no.~2,
  357--387 (en).

\bibitem[Stu95]{sturmfels_grobner_1995}
Bernd Sturmfels, \emph{Gröbner {Bases} and {Convex} {Polytopes}}, University
  {Lecture} {Series}, vol.~8, American Mathematical Society, December 1995
  (en-US).

\bibitem[Stu03]{sturmfels_algebraic_2003}
\bysame, \emph{Algebraic {Recipes} for {Integer} {Programming}},
  arXiv:math/0310194 (2003), arXiv: math/0310194.

\bibitem[TTN95]{tayur_algebraic_1995}
Sridhar~R. Tayur, Rekha~R. Thomas, and N.~R. Natraj, \emph{An algebraic
  geometry algorithm for scheduling in presence of setups and correlated
  demands}, Mathematical Programming \textbf{69} (1995), no.~1-3, 369--401
  (en).

\bibitem[VK18]{venturelli_reverse_2018}
Davide Venturelli and Alexei Kondratyev, \emph{Reverse {Quantum} {Annealing}
  {Approach} to {Portfolio} {Optimization} {Problems}}, arXiv:1810.08584
  [quant-ph, q-fin] (2018), arXiv: 1810.08584.

\end{thebibliography}

\appendix
\newpage
\section{Graver Basis via Classical Methods}
\label{sec:classicalgraver}

Classically any algorithm that can calculate all inclusion-minimal elements of any sublattice $\mathcal{L} \subseteq {\mathbb{Z}^n}$ can be used to calculate Graver basis.% if sublattice is $\mathcal{L} = \ker \mathbb{N}(A)$.
All such algorithms have a computational complexity that is exponential.

~~\\
%\subsection{Pottier's Algorithm}
 Pottier's algorithm \cite{pottier_euclidean_1996} -- also called \textit{completion procedure} - starts with any lattice basis $\mathcal G$  of $\mathcal{L}$, and iteratively adds new vectors to $\mathcal G$ until any vector $z \in \mathcal{L}$  can be written as a positive linear sign-compatible (same orthant) representation:

\begin{equation}
\begin{array}{*{20}{c}}
  {z = \sum\limits_i {{\lambda _i}} {g_i}}&{{\lambda _i} \in {\mathbb{Z}_{ > 0}}}&{{g_i} \in \mathcal G}&{{g_i} \sqsubseteq z} 
\end{array}
\end{equation}
If $\mathcal G$ reaches the representation property, then it contains all  $\sqsubseteq $-minimal elements. 
%For Pottier's algorithm, we need to use  normal form algorithm in Algebraic Geometry. 
The vector version is shown in Algorithm \ref{alg:NF}.
~~\\
{\centering
\begin{minipage}{1.0\linewidth}
\begin{algorithm}[H]
\caption{Normal Form}\label{alg:NF}
\small{
\begin{algorithmic}[1]
\STATE \textbf{input} vector $s \in \mathcal{L}$, set $\mathcal G \subset \mathcal{L}$
\STATE \textbf{output} vector $r = normalForm(s,G) \in \mathcal{L}$ s.t. $s = \sum\limits_i {{\alpha _i}} {g_i} + r$, ${\alpha _i} \in {\mathbb{Z}_{ > 0}}$, ${g_i}, r \sqsubseteq s$
\STATE Initialize: $r \leftarrow s$
\WHILE{$\begin{array}{*{20}{c}}
  {\exists g \in \mathcal G}&{s.t.}&{g \sqsubseteq r} \end{array}$}
\STATE $r \leftarrow r - g$
\ENDWHILE
\RETURN $r$
\end{algorithmic}
}
\end{algorithm}
\end{minipage}
}
~~\\
When $r$ becomes zero, $s$ is integer linear combination of ${g_i}$’s in the same orthant. If $r$ does not reach zero, the last $r$ has to be added to $\mathcal G$ thus in future iterations $r$ becomes zero.
This is shown in Algorithm \ref{alg:POT}.
~~\\
{\centering
\begin{minipage}{1.0\linewidth}
\begin{algorithm}[H]
\caption{Pottier}\label{alg:POT}
\small{
\begin{algorithmic}[1]
\STATE \textbf{input} Generating initial set $F \subseteq \mathcal{L} = \ker \mathbb{N}(A)$
\STATE \textbf{output} Graver basis set $G \subseteq \mathcal{L}\backslash \left\{ {\mathbf{0}} \right\}$
\STATE Initialize symmetric set: $G\leftarrow F \cup ( - F)$
\STATE Generate S vector set: $C \leftarrow \bigcup\limits_{f,g \in \mathcal G} {\left\{ {f + g} \right\}}$
\WHILE {$C \ne \emptyset$}
\STATE $\begin{array}{*{20}{c}}
  {\forall s \in C}&:&{r \leftarrow normalForm(s, \mathcal G)}&{and}&{C \leftarrow C\backslash \left\{ s \right\}} 
\end{array}$
\IF{$r \ne 0$}
\STATE Update: $\begin{array}{*{20}{c}}
  {\mathcal G \leftarrow  \mathcal G \cup \left\{ r \right\}}&{and} &{C \leftarrow C \cup \left\{ {r + {g_i}} \right\}}&{{g_i} \in \mathcal G} 
\end{array}$
\ENDIF
\ENDWHILE
\RETURN $\mathcal G$
\end{algorithmic}
}
\end{algorithm}
\end{minipage}
}
~~\\
It can be shown \cite{de_loera_algebraic_2012} that the above algorithm terminates with all inclusion minimal elements of $\mathcal{L}$. (There remain non inclusion minimal terms that should be filtered out  based on the inclusion minimal concept.)  Other algorithms for computing Graver bases exist such as project and lift
\cite{cornuejols_decomposition_1997, hemmecke_computation_2002} and using  Lawrence lifting \cite{sturmfels_grobner_1995}.

\section{Quantum Annealing and D-Wave Processors}
\label{sec:quantumannealer}
 
\subsection{Quantum Annealing}

~~\\
Quantum annealing \cite{farhi_quantum_2000,mcgeoch_adiabatic_2014} uses
the quantum adiabatic evolution  \cite{kato_adiabatic_1950} to solve optimization problems (QUBOs). (See \cite{cmumt} for a Morse theoretical description for the quantum adiabatic evolution.) 
 %Quantum annealing, along with the D-Wave processor, has been the focus of much research. We refer the interested reader to references .\\
%The adiabatic evolution in quantum mechanics inspires the quantum annealing. Based on adiabatic evolution
This is done by slowly evolving the ground state of some known system into the sought ground state of the problem Hamiltonian. For instance, D-Wave 2000Q  \cite{harris_experimental_2010,johnson_quantum_2011,bunyk_architectural_2014} implements quantum annealing using the time
dependent Hamiltonian
\begin{equation}
    \mathcal{H}(s) = A(s){\mathcal{H}_I} + B(s){\mathcal{H}_P}
\end{equation}
%if the changes in system are slow enough so there stays a gap between the ground state and the rest of the system’s energy states . In quantum annealing they specify an initial  Hamiltonian ${\mathcal{H}_I}$, whose ground state is known, and a problem Hamiltonian ${\mathcal{H}_P}$, whose ground state depicts the solutions to the problem in hand \cite{farhi_quantum_2000}. Then they interpolate from initial Hamiltonian to problem Hamiltonian by generating a merged time-evolving Hamiltonian $\mathcal{H}(s) = A(s){\mathcal{H}_I} + B(s){\mathcal{H}_P}$,
where $A(s)$ is monotonically decreasing while $B(s)$ is monotonically increasing with respect to normalized (slow) time $s = \frac{t}{\tau} \in \left[ {0,1} \right]$.  $\tau$ is  the total annealing time. The initial Hamiltonian (with known ground state) is a transverse magnetic field ${\mathcal{H}_I} =  - \sum\limits_i {\sigma _i^x}$ where $\sigma _i^x$ is the  ${i^{th}}$ Pauli $x$-matrix.
The problem Hamiltonian ${\mathcal{H}_P}$ is  \cite{barahona_computational_1982}:
\begin{equation}
{\mathcal{H}_P} = \sum\limits_{i \in V} {{h_i}} \sigma _i^z + \sum\limits_{(i,j) \in E} {{J_{ij}}} \sigma _i^z\sigma _j^z,
\end{equation}
where the parameters ${h_i}$ and ${J_{ij}}$ encode the particular problem instance. The $\sigma _i^z$ are Pauli $z$-matrices. 
A measurement of the final state (i.e.,end of the adiabatic evolution at $s=1$) will yield a solution of the problem.  
%{\color{red} Cite our Morse paper: it gives a good summary of the adiabatic evolution (adiabatic thm for all type of spec, avoid crossing, historical notes) and gives a new topological description using the language of Morse theory -- \cite{cmumt}. }  
\subsection{Quantum Annealer from D-Wave}
%Like other quantum computation paradigms, quantum-annealing concept in physical realization suffers from magnetic noise, thermal excitations due to temperature and lack of enough coherence \cite{childs_robustness_2001,amin_decoherence_2009,albash_decoherence_2015}. Realization issues, the inherent nature of implemented initial Hamiltonian and specific D-Wave qubit circuit limitations, has caused a heated debate and research over whether D-Wave processors deliver any speedup in general \cite{shin_how_2014,muthukrishnan_tunneling_2016,king_quantum_2017,ronnow_defining_2014,katzgraber_glassy_2014,boixo_experimental_2013}. In this work we avoid the speedup concern and use D-Wave processor as a physical quantum annealer to evaluate the performance and competence of our proposed hybrid nonlinear integer-programming algorithm. This does not necessarily mean that we ignore the potentials and deficiencies of D-Wave processor.\\
We had access to a D-Wave Quantum Annealer 
2000Q\textsuperscript{TM} (C16-VFYC solver) administered by NASA Quantum Artificial Intelligence Laboratory (QuAIL). This processor operates at $20 (\pm 5) milliKelvin$ and designed with 2048 qubits with a 95.5 qubits yield in its 16 by 16 block \textit{Chimera} configuration. The \textit{Chimera} structure has a limited connectivity as shown in Figure \ref{fig:chimera}.
%Generally couplers cannot be manufactured connecting any two qubits which are far spatially from each other, due to surface layout limitations. 
Therefore, the graph of a specific QUBO problem needs to be embedded into the graph of hardware~(Figure~\ref{fig:chimera}).  Embedding an input graph into the hardware graph  is a specific instance of a graph homomorphism and its associated decision problem is in general NP-complete \cite{choi_minor-embedding_2008, cmuag}. 
\begin{figure}[h]
\centering
\includegraphics[width=7cm]{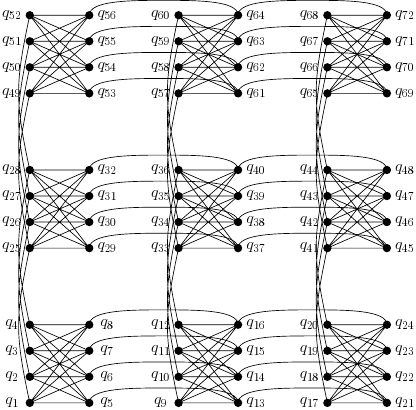}
\caption{Schematic representation of a 3 by 3 block \textit{Chimera}}
\label{fig:chimera}
\end{figure}

\subsubsection{D-Wave SAPI}
D-Wave's software application programming interface (SAPI) is an application layer built to provide resource discovery, permissions, and scheduling for quantum annealing. The codes for this project are implemented using Matlab SAPI (version 3.0). For embedding the problem graph into hardware graph and also for parameter settings, the sapiFindEmbedding (default parameters) and sapiEmbedProblem (adaptive problem per chain strength) software modules are used. To solve, sapiSolveIsing and sapiUnembedAnswer modules (with two distinct chainbreak strategies discussed in  4.3.1) are used.  The maximum number of reads available ($10000$) is used to increase the efficiency of each call (the number of unique optimal and nonoptimal solutions).

\subsubsection{Chainbreak strategy}\label{subsec:chainbreak}
Embedding a problem graph into the hardware graph has two parts: topological and problem value mapping. The second aspect deals with setting of the parameters.  All qubit values in a \textit{chain} are expected to remain at the same value (+1 or -1) at the end of quantum annealing process. Due to situations in which the chain couplers ($J_{F}$, \textit{identified} couplers) are not strong enough in comparison with problem couplers, chains frequently break.
To recover from these breaks, D-Wave has provided two post-processing strategies:
\begin{itemize}
    \item \textit{Minimize energy}, which is the default choice. This approach walks each broken solution down to a local minimum by repeatedly flipping random bits to reduce solution cost, somewhat like a zero-temperature simulated annealing, but without moves that do not change the energy (\cite{venturelli_reverse_2018}). 

\item \textit{Majority vote}, on the other hand, is computationally less expensive: it sets the chain value according to a \textit{majority vote} of qubits in the chain, and a random value in case of ties.
\end{itemize}
In structured and dense problems, where the embedding is close to a complete graph embedding, all variables end up having long chains thus the probability of chain break in them increases. In such situations, the effect of chainbreak post processing might exceed the effect of quantum annealing itself. Therefore, in the results, we report the chain-break strategy used as well as the percentage of chains broken. This indicates how much of the problem has been impacted by chainbreak post-processing. 

\subsubsection{Annealing time}

Annealing time ($\tau$) is the (a) forward (or reverse) dynamic time from the start of annealing to the end, or (b) from the start to the pause, or (c) from end of pause to end of annealing.
Our experimental tests show that when we select a large annealing time(i.e. $290 \mu sec$), we have fewer chain breaks, but also fewer unique solutions. When we chose a small annealing time (i.e. $1-10 \mu sec$) we have more chain breaks, but higher number of unique solutions (using minimize energy chain-break strategy). 
% It is not clear if one choice of annealing time is better than the other uniformly.

~~\\
\textit{Pause} is the static stopping time (usually in the middle of forward or reverse annealing) that allows local exploration through thermal hopping and enhances the analog of classical gradient descent. The pause time ($\rho$) is usually larger than forward (or reverse) annealing time ($\tau$). 
In \textit{reverse annealing}, the system is initialized with $B(1) = 1$ 
and $A(0) =0$, with all used qubits preset to a classically known solution. The annealing then turns back to somewhere in the middle of the schedule, where the initial (transverse field) Hamiltonian and problem Hamiltonian have roughly the same strength, and then pauses. After the pause, there is a third phase, which is the final forward annealing. The reasoning behind reverse annealing is that by starting from a known local minimum of the problem, the reciprocation of quantum and thermal fluctuations might help the quantum state to \textit{tunnel} through the energy barriers (in reverse phase), while the pause and the (final) forward annealing allows the system to relax and thermalize in vicinity of a better minimum. In our results, we report the annealing schedule, because it is an important factor in the quality and quantity of the results. 

\end{document}